\documentclass[floatfix,useAMS,usenatbib]{mnras}
\usepackage[greek,british]{babel}
\usepackage{amsmath}
\usepackage{graphicx,epsfig}
\usepackage{multirow}
\usepackage{verbatim}
\usepackage[figuresright]{rotating}
\usepackage{txfonts}
\usepackage{times}
\usepackage[T1]{fontenc}
\usepackage{aecompl}

\title[``Ears'' formation in Supernova Remnants]
     {``Ears'' formation in Supernova Remnants:  Overhearing an interaction history with bipolar circumstellar structures}

\author[Chiotellis,  Boumis \& Spetsieri]
  {A. Chiotellis,$^1$\thanks{a.chiotellis@noa.gr}
  P. Boumis,$^1$ Z. T. Spetsieri$^1$\\  
  $^1$Institute for Astronomy, Astrophysics, Space Applications
and Remote Sensing, National Observatory of Athens,
15236, Penteli, Greece}

\date{Accepted 2020 November 10; Received 2020 November 8; in original form 2020 October 29}

\pubyear{2020}

\begin{document}
\pagerange{\pageref{firstpage}--\pageref{lastpage}}
\maketitle
\label{firstpage}

\begin{abstract} 

A characteristic feature that is frequently met in nearby supernova remnants (SNRs) is the existence of two antisymmetric, local protrusions that are projected as two ``ears'' in the morphology of the nebula. In this work, we present a novel scenario for the ``ear'' formation process according to which the two lobes are formed through the interaction of the SNR with a bipolar circumstellar medium (CSM) that was surrounding the explosion center. We conduct two dimensional hydrodynamic simulations and we show that the SNR shock breakout from the bipolar CSM triggers the inflation of two opposite protrusions at the equator of the remnant that retain their size and shape for several hundreds up to a few thousand years of the SNR evolution. We run a set of models by varying the supernova (SN) and CSM properties and we demonstrate that the extracted results reveal a good agreement with the observables, regarding the ``ears'' sizes, lifespan, morphology and kinematics. We discuss the plausibility of our model in nature and we suggest that the most likely progenitors of the "ear-carrying" SNRs are the Luminous Blue Variables or the Red/Yellow Supergiants for the SNRs resulted by core collapse SN events and the symbiotic binaries or the planetary nebulae for the remnants formed by Type Ia SNe. Finally, we compare our model with other ``ear'' formation models of the literature and we show that there are distinctive differences among them, concerning the ``ears'' orientation and the phase in which the ``ear'' formation process occurs.

\end{abstract}


\section{Introduction}\label{sec:Intro}

Supernova remnants (SNRs) are the aftermath of supernova (SN) explosions that result from the interaction of the supersonically moving stellar ejecta with the ambient medium. These celestial nebulae reveal complex morphological properties in several spacial scales and all over the electromagnetic spectrum. There is a consensus that the complex properties of SNRs are mainly determined by two parameters: a) the nature of the parent stellar explosion and b) the interaction of the SN ejecta with circumstellar structures sculpted by the mass outflows of their progenitor stellar systems. Thus, peculiar morphological features of SNRs (e.g. SNR's asymmetries and inhomogeneities, hydrodynamic instabilities, rings, bow shocks, jets) host crucial encoded information about the SN explosion mechanism and the nature and evolution of their stellar progenitors. In order to decipher this information and meticulously separate the `cause and effect' that lead to the observed SNRs properties detailed modeling is required \citep[e.g.][]{Orlando2020,Warren2013, Chiotellis12, Badenes2006, Ellison2004,  Dwarkadas1998}.

\begin{figure}
\includegraphics[trim=0 0 0 0 0,clip=true,width=\columnwidth,angle=0]{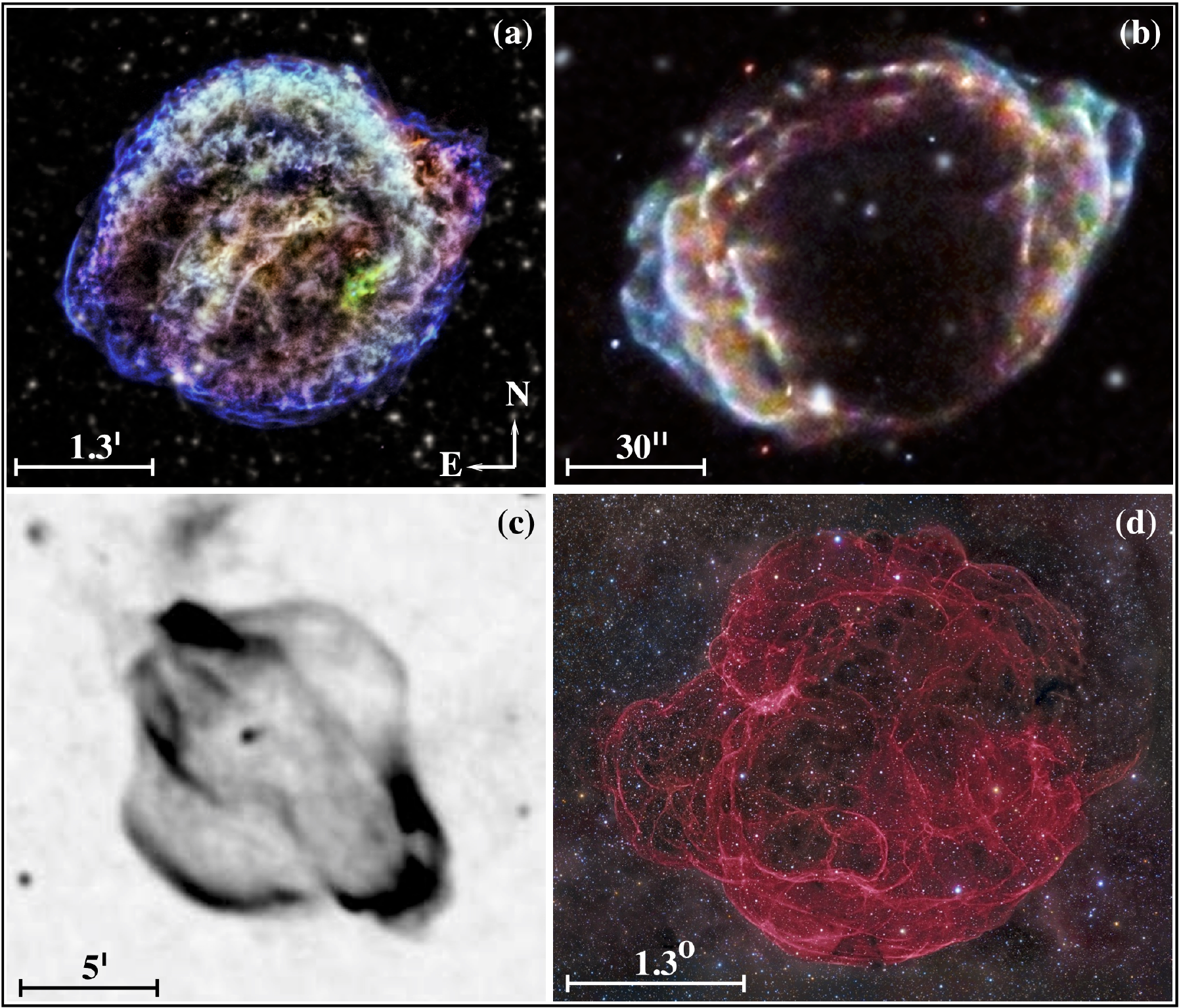}
\caption {Characteristic cases of SNRs that reveal two antisymmetric ``ears'' in their morphology: (a) The X-ray image of Kepler's SN \citep[SN1604;][]{Reynolds2007}, (b) the X-ray image of the young SNR G1.9+0.3 \citep{Borkowski2013}, (c) the radio image of G309.2-06 at 1.3 GHz \citep{Gaensler1998} and (d) the $\rm H_{\alpha}$ image mosaic of S147 \citep{Drew2005}. North (N) is to the top and East (E) to the left in all images (as shown in (a)).}
\label{fig:SNRs_ears}
\end{figure}

 A morphological peculiarity that is frequently met in SNRs is the presence of two antisymmetric protrusions at the outermost region of the remnant. These protrusions, which are frequently termed as ``ears'', penetrate and deform the forward shock of the SNRs shaping two opposite, local lobes in the overall morphology of the remnant. The ``ears'' have been observed in all types of SNRs (Type Ia and Core Collapse) and evolutionary stages (from young X-ray bright remnants till well evolved non-adiabatic SNRs). Characteristic cases are the Galactic SNRs of Kepler's (SN 1604), G1.9+0.3, G309.2-06 and S147 (Fig.\ref{fig:SNRs_ears}; see also \citet{Bear2017} and \citet{Tsebrenko2015b} for a complete list of SNRs that possess antisymmetric ``ear-like'' features). A common characteristic for all cases is the remarkable symmetry of the ``ears'' in terms of brightness and shape, as well as, their opposed positions in respect to the center of the remnant. These properties advocate an axis or central symmetric formation mechanism more relevant to the nature of their parent stellar progenitor and/or SN explosion than local interstellar medium (ISM) inhomogeneities.

Up to date, the formation of the ``ears'' in SNRs has been attributed to the launch of two opposite jets that accompany the SN explosion, or are triggered  after it, which protrude the forward shock of the SNR and inflate the two opposite lobes \citep{Millas2019, Yu2018, Bear2017, Grichener2017,Tsebrenko2013, Castelletti2006, Gaensler1998}. For the case of SNRs resulted by Type Ia SNe (SNe Ia), it has been additionally suggested that the ``ears'' pre-existed in the morphology of a Planetary Nebula (PN) that surrounded the explosion center \citep{Tsebrenko2013, Tsebrenko2015}. In this scenario the ``ears'' of the SNR were sculptured by the interaction of the SN ejecta with the shell of the surrounding ``ear-carrying'' PN . Finally, \citet{Tsebrenko2015c} suggested that the ``ears'' of SNRs  resulted by SNe Ia, are shaped by iron clumps, or `bullets' in the dense ejecta, formed along a common axis due to the rotation of the white dwarf  progenitor \citep[see also][for a thorough discussion on the evidence and constrains imposed by the ``ears'' in SNRs regarding their parent SNe Ia explosion]{Soker2019}. 

\citet{Blondin1996} suggested an alternative scenario on the formation of the SNR's protrusions, according to which the two ``ears'' are inflated during the evolution of the remnant within an axisymmetric circumstellar structure characterised by a high density enhancement at the equatorial plane. The authors considered a circumstellar medium (CSM) described by a wind bubble in which the density is a function of the polar angle and increases from the poles to the equator of the system. Performing hydrodynamic simulations they showed that the interaction of the SN ejecta with such an axisymmetric ambient medium results to the appearance of two protrusions close to the polar axis of the remnant that -depending on the polar density gradient- can be extended at a length up to 2-4 times the overall radius of the remnant.

In this study, we propose a novel mechanism for the ``ears'' formation in SNRs. We retain the idea of \citet{Blondin1996} about the SNR interaction with a bipolar and equatorially confined circumstellar structure but we examine the ``ear'' formation process under a different aspect. We state that the two opposite protrusions observed in several SNRs are formed at the equatorial plane of the remnant during the forward shock's breakout from the surrounding bipolar CSM. These protrusions maintain their shape from hundreds up to a few thousand years after the shock breakout giving the impression of two ``ears'' in the overall morphology of the SNR. In other words, we claim that the SNRs that host two opposite lobes had an interaction history with a dense and equatorially confined CSM and are currently evolving in a less dense ambient medium. We demonstrate using hydrodynamic simulations that this model can account for the morphological and kinematic properties of the antisymmetric protrusions observed in several SNRs.

The paper is organised as follows: In Section \ref{sec:Hydro}, we describe our numerical model for the SNR-bipolar CSM interaction and we present the results of our hydrodynamic simulations. Moreover, we emphasise on the ``ear'' formation mechanism  and we articulate the dependency of our model's results on the properties of the CSM and the SN explosion.  In Section \ref{sec:Discuss}, we compare the results extracted by our models to the relevant observable, we discuss the plausibility of our model in nature and finally, we present the differences between our models and the rest the ``ear-formation models'' presented in the literature. Finally, in Section \ref{sec:Summary}, we summarize our main results and conclusions.


\section{Hydrodynamic modeling }\label{sec:Hydro}

The simulations of the bipolar CSM formation and the subsequent interaction of the SNR with it are performed using the hydrodynamic code AMRVAC \citep{Keppens03}. We employ a two dimensional (2D) grid in spherical coordinates and assume symmetry in the third dimension of the azimuthal angle. The radial span ($R$) of the computational domain is 18~pc  while the polar angle ($ \theta$) ranges from $0^{\rm o}$ to $180^{\rm o}$. Our grid consists of $(R \times \theta)= 360 \times 120 $ grid cells. We exploit the adaptive mesh capabilities of the AMRVAC code by using four refinement levels of resolution, at each of which the resolution is doubled as a result of large gradients in density and/or energy. Hence, the maximum effective resolution becomes $6.25 \times 10^{-3}$~pc by $0.19^{\rm o}$. Radiative cooling is prescribed using the cooling curve of \citet{Schure09}.


\subsection{The formation of the bipolar CSM }\label{subsec:CSM_hydro}

\begin{figure}
\includegraphics[trim= 95 45 85 45,clip=true,width=\columnwidth,angle=0]{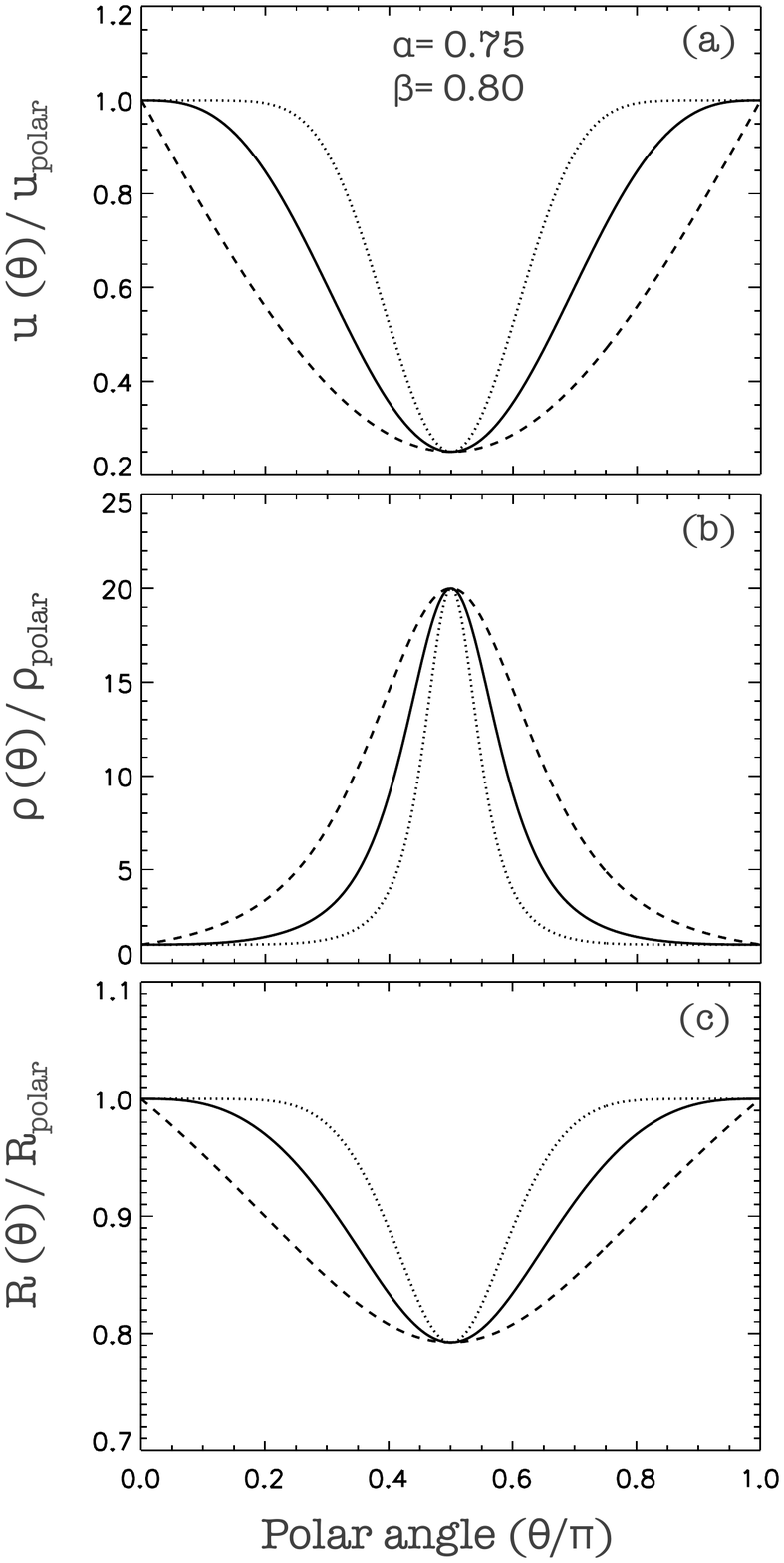} 
\caption { The bipolar distribution of the CSM velocity (a), density (b) and outer radius (c), normalized over their polar values. The graphs are described by Eq.  \ref{eq:u_theta}, \ref{eq:rho_theta} and \ref{eq:radii_ratio}, assuming $\alpha= 0.75$ and $\beta= 0.8$. The solid lines correspond to $k= 3$, while the dashed and dotted lines to $k= 1$ and $k= 6$, respectively. }
  \label{fig:bip_ratios}
\end{figure}

\begin{figure}
\includegraphics[clip=true,width=\columnwidth,angle=0]{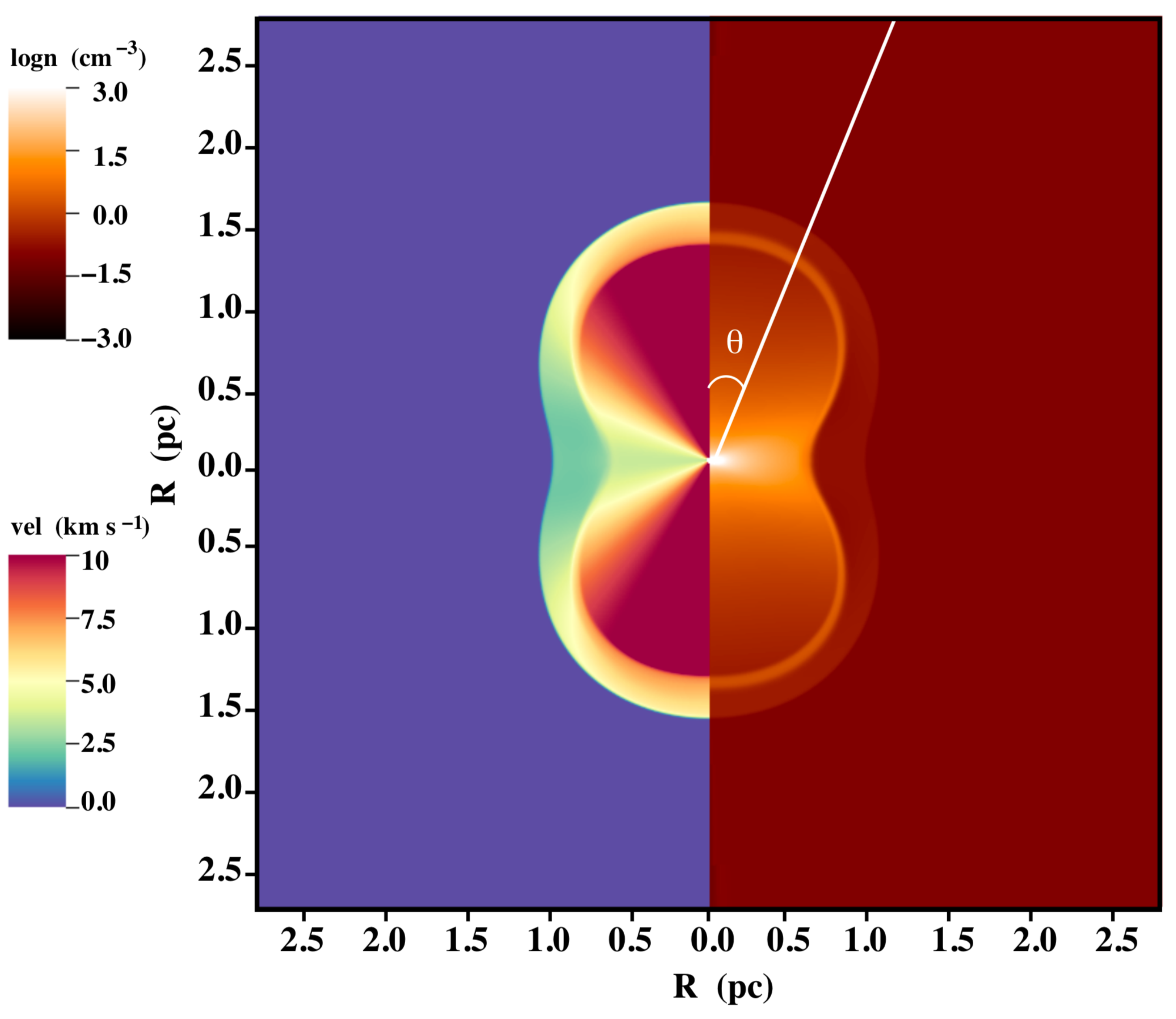} 
\caption {The 2D profile of a bipolar circumstellar structure. The stellar wind emanates from the axis origin where we impose an equatorially confined flow  to enter the grid. The right plot shows the 2D density profile while the left displays the velocity distribution of the circumstellar structure. The used wind and ISM parameters are summarised in Table \ref{tab:ModelA}. }
  \label{fig:CSM_sample}
\end{figure}

We first simulate the formation of the bipolar circumstellar structure. Considering that the mass-losing progenitor star/system is located at the axis origin of our grid, the CSM bubble is created by injecting material into the computational domain through the cells at the inner radial boundary. This inflow is in the form of a slow, continuous and equatorially focused stellar wind. Assuming that the stellar wind is axisymmetric in the azimuthal dimension we describe the properties of the bipolar wind with the following trigonometrical functions:

\begin{equation} \label{eq:u_theta}
u_{w}(\theta)= u_{w, p} \left[1 - \alpha \mid \sin \theta \ \mid ^{k}  \right]
\end{equation}

\begin{equation} \label{eq:rho_theta}
 \rho(\theta)= \frac {\dot{M}_{p} \left[1 - \beta \mid \sin \theta \ \mid ^{k}  \right]^{-1}} {4 \pi r^2 u_{w}(\theta)}
\end{equation}
where $\rho(\theta)$, $u_{w}(\theta)$ the wind density profile and  terminal velocity at the polar angle $\theta$, respectively.  $u_{w, p}$ is the  terminal velocity and $\dot{M}_{p}$ the mass loss rate of the stellar wind at the poles of the system ($\theta=0^{\rm o}$). In Eq. \ref{eq:rho_theta} the symbol $r$ refers to the radial distance from the mass losing star. The $\alpha, \beta$ and $k$ are constants which determine the polar distribution of the CSM density and velocity from the poles to the equatorial plane. Specifically, $ \alpha $ and $ \beta $ ($ 0 \leq \alpha, \beta < 1 $) stand for the ratio of the polar and equatorial velocity and density, respectively where for a given $r$ we get: $\rho_{eq}/\rho_{p}= [(1-\alpha)(1-\beta)]^{-1}$ and $u_{w,eq}/u_{w,p}=(1-\alpha)$. Finally, the index $k > 0 $ determines the angular density gradient and the confinement level of the stellar wind at the equatorial plane. For small values of $k$ the CSM density and velocity gradually change (increases and decreases, respectively) from poles to equator, while for $k >> 1$ a disk like morphology is shaped, confined in the equatorial plane that is surrounded by a roughly spherical  wind bubble (see Fig. \ref{fig:bip_ratios}a,b).

The polar distribution of the CSM density and radial velocity has a direct impact on the radius of the wind bubble at each polar angle, which in turn determines the overall shape of the circumstellar structure. Following \citet{Weaver1977}, the outer radius of an adiabetically expanding wind bubble ($R_b$) is proportional to: $R_b \propto (L_w/n_{ism})^{1/5} t^{3/5} $, where $Lw$ the wind mechanical luminosity ($Lw= 1/2 \dot{M} u_w^2$), $n_{ism}$ the ISM density and $t$ the age of the bubble. Thus, the ratio of the wind bubble radii in two different polar angles ($\theta_1$ and $\theta_2$) at a given time is:  

\begin{equation} \label{eq:radii_ratio}
 \frac{R_b (\theta_1)}{R_b (\theta_2)}= \left(\frac {\dot{M} (\theta_1)} {\dot{M} (\theta_2)}\right)^{1/5} \left(\frac {u_w (\theta_1)} {u_w (\theta_2)}\right)^{2/5}
\end{equation}
\\
Applying in Eq. \ref{eq:radii_ratio} the adopted trigonometrical descriptions of the wind properties (Eq.\ref{eq:u_theta} and \ref{eq:rho_theta}) the ratio of the wind bubble radius at the equator over the one at the poles is given by: 

\begin{equation} \label{eq:radii_ratio_ab}
 \frac{R_{b,eq}}{R_{b,pol}}=\left(1 - \alpha \right)^{2/5} \left( 1- \beta \right)^{-1/5} 
\end{equation}
\\
Hence, according to Eq. \ref{eq:radii_ratio_ab} as the constant $\alpha$ increases and $\beta$ decreases the equatorial dense waist of the CSM becomes narrower with respect to the overall size of the bubble (see Fig. \ref{fig:bip_ratios}c).

\begin{table}
\centering

\caption{The  wind and ISM parameters adopted for the simulation of the  bipolar CSM  illustrated in Fig. \ref{fig:CSM_sample}. Here $n_{ism}$ and $T_{ism}$ is the density and temperature of the ISM, while $\dot{M}_p$ and $u_{w,p}$ denote the wind mass loss rate and terminal velocity at the poles, respectively. $T_w$ and $\tau_w$ stand for the wind's temperature and the time interval of the wind phase and finally, $\alpha$, $\beta$, $k$ are the constants used in the trigonometrical functions (Eq. \ref{eq:u_theta}, \ref{eq:rho_theta}).    }
\label{tab:ModelA}
\begin{tabular}{cc|cc }
\hline
\multicolumn{4}{|c|}{Bipolar Wind Properties} \\
\hline
$n_{ism}~\rm{(cm^{-3})}$              &    0.1     & $\alpha$ &  0.75  \\   
$T_{ism}~\rm{(K)}$                    &   1000     & $\beta$  &  0.8    \\
$ \dot{M}_p~ \rm(M_{\odot}~yr^{-1})$  &  $2 \times 10^{-6}$ &  k       &    3    \\
$ u_{w,p}~\rm(km~s^{-1})$             & 10 &$u_{w,eq}/u_{w,p}$ & 0.25  \\
$T_{w}~\rm{(K)}$                      & 1000 &$\rho_{eq}/\rho_{p}$ &  20  \\
$\tau_w ~\rm{(Myr)}$                  &0.15  & $R_{b,eq}/R_{b,p}$ & 0.79 \\       
\hline
\end{tabular}
\end{table}

\begin{figure*}
\includegraphics[trim=0 0 0 0 0,clip=true,width=\textwidth,angle=0]{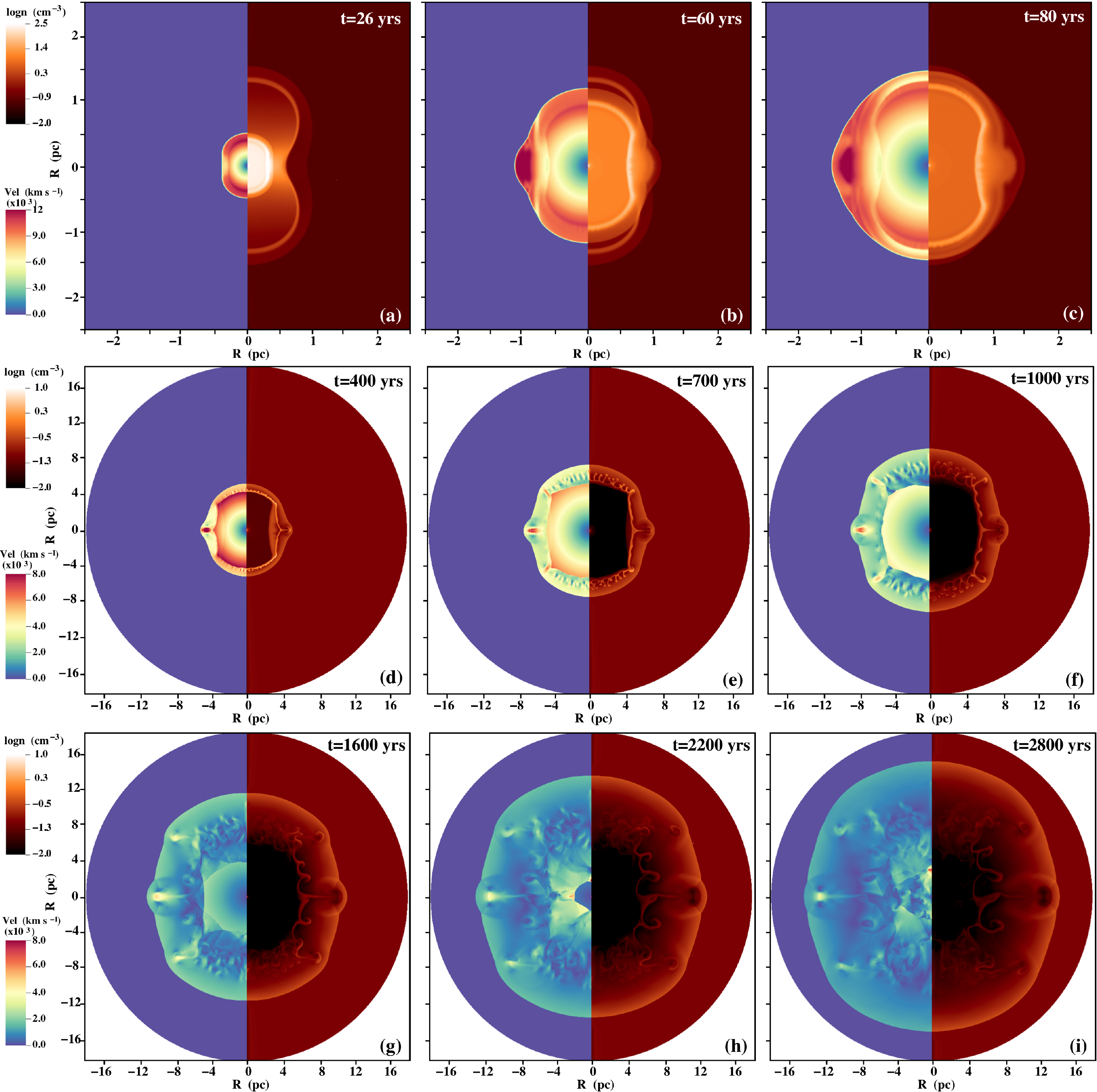} 
\caption{The evolution of the SNR as resulted by the interaction of the SN ejecta with the bipolar circumstellar structure of Fig. \ref{fig:CSM_sample}.  The right part of each plot depicts the 2D density contours of the SNR while the left one the corresponding gas velocity. The time indicates the age of the SNR at each snapshot. Note that both the axis and color scale in the first row are different to the other two.}
\label{fig:SNR_evol}
\end{figure*}

Figure \ref{fig:CSM_sample} illustrates a typical bipolar circumstellar structure as resulted by our simulations. The used parameters of the equatorially focused stellar wind are tabulated in Table \ref{tab:ModelA}. The resulted circumstellar structure reveals an hourglass morphology, forming an equatorial disk of dense, slow moving material. From inside out the four main regions of the wind bubble are clearly depicted: a) the freely expanding wind where $\rho_{wind} \propto r^{-2}$, b) the shocked wind shell, c) the shell of shock ambient medium and d) the outermost region of unperturbed ISM. The inner density jump corresponds to the position of the termination shock, while the outer one to this of the forward shock. The equatorial to polar density and velocity ratios of the CSM are $\rho_{eq}/\rho_{pol}= 20.0$ and $u_{w,eq}/u_{w,pol}= 0.25$, respectively. Finally, the extracted relative ratio of the structure's outer radii in the poles and equator is $R_{b,eq}/R_{b,pol}= 0.7$, $\sim 10 \%$ smaller than the prediction of the analytical approach (cf. Eq.~\ref{eq:radii_ratio_ab}). This difference is attributed to  radiation cooling that occurs in the dense equatorial region of the bipolar CSM.

 
\begin{figure*}
\includegraphics[clip=true,width=\textwidth,angle=0]{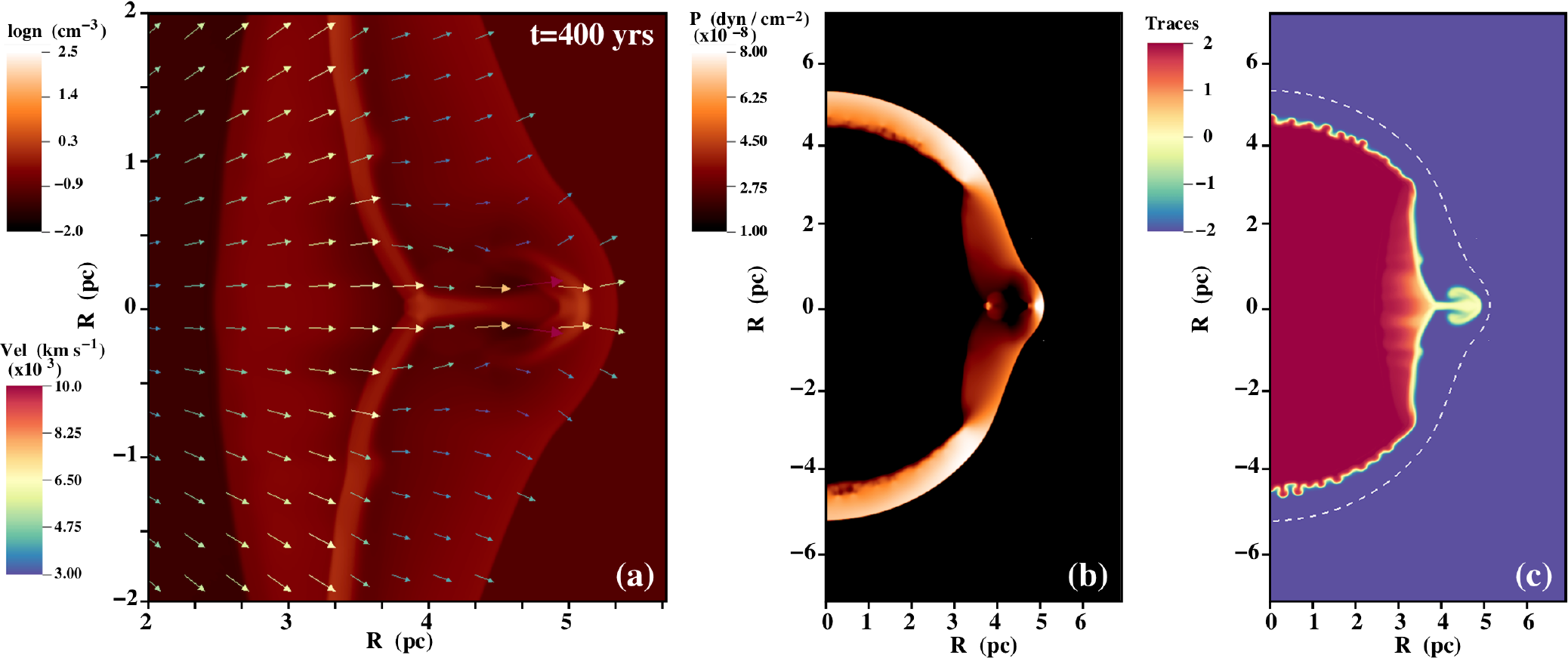} 
\caption { (a): The same snapshot of the SNR  as in Fig. \ref{fig:SNR_evol}d but zoomed in the region where the ``ear'' is formed. In addition, the velocity vectors of the flow have been added. (b): The 2D  gas pressure contours of the SNR at the same age. (c): The traces of three components that constitute the SNR: the SN ejecta, the circumstellar material and the interstellar medium. The values close to two indicate pure SN ejecta material, these around zero refer to regions where the CSM material dominates and finally, these close to minus two to regions of the grid of purely interstellar material. The values between these numbers correspond to regions where mixing among the three components occurs. The dashed line indicates the position of the SNR forward shock in this specific snapshot. }
\label{fig:ear_formation}
\end{figure*}

\subsection{The interaction of the SNR with the bipolar CSM and the ``ears'' formation }\label{subsec:SNR_hydro}

To model the interaction of a SNR with a bipolar CSM, we introduce the supernova ejecta in the center of the circumstellar structure depicted in Fig. \ref{fig:CSM_sample}, and we let the SNR to evolve and interact with the surrounding medium. The adopted energy of the SN is $10^{51}$~erg, while the ejecta mass $1.38~\rm M{_{\odot}}$. Finally, following \citet{Truelove99}, the ejecta density profile is described by a constant density core with an envelope that follows a power law of $\rho \propto r^{-n_{\rm SN}}$ with $n_{\rm SN} = 7$ , while the ejecta's velocity increases linearly \footnote{The adopted SN properties correspond to a normal, Chandrasekhar-mass SN Ia explosion. In Sect. \ref{subsec:models_comp}, we extend our model for cases more relevant to core collapse SNe.}.

Figure \ref{fig:SNR_evol} shows the 2D density and velocity contours for a sequence of snapshots of the SNR evolution, as resulted by our hydrodynamic simulations.   

At the initial phase of the SNR evolution, the remnant is expanding within the circumstellar bubble (Fig. \ref{fig:SNR_evol}a). At the equatorial plane of the SNR, the forward shock encounters denser material than the rest of the remnant and thus, it expands slower. Consequently, the SNR starts to depart from the spherical symmetry and it reveals a bipolar morphology. 

Around 60 years after the SN explosion (Fig. \ref{fig:SNR_evol}b), the forward shock reaches the outermost region of the equatorial disk and starts to propagate at the surrounding ISM. The shock breakout from the CSM is accompanied by a significant acceleration of the forward shock at this portion of the SNR, due to the low ISM density and the high post shock pressure. The SNR is expanding in the equatorial region with a velocity of $\simeq 1.5 \times 10^4 ~ \rm {km~s^{-1}}$, while the rest of the remnant that remain within the bubble has an expansion velocity of $\simeq 9 \times 10^3 ~ \rm {km~s^{-1}}$. As a result, a high velocity protrusion is formed in the equatorial region of the SNR. 

After 80 yrs of evolution, the SNR has entirely swept up the CSM and starts to expand into the homogeneous ISM (Fig. \ref{fig:SNR_evol}c). The SNR's reverse shock carries on its morphology signatures of the interaction history with the bipolar CSM, being more evolved in the equatorial plane of the SNR. The forward shock is getting accelerated and starts to establish a more spherical shape. Nevertheless, at the equator of the remnant the initially formed protrusion retains its high velocity and forces the remnant to shape a local bulge.  
 
As the SNR progresses further, the shocked gas behind the formed shock starts to decelerate and the remnant's contact discontinuity is  subject to Rayleigh-Taylor (RT) instability (Fig. \ref{fig:SNR_evol}d-f). The formed RT fingers after their initial growth, reach saturation and subsequently they get deformed, bend and finally fall back. The maximum size they reach is about half the width of the shocked shell and thus, they do not perturb the forward shock during the whole SNR evolution \citep[see also][]{Chevalier1992}. An intriguing exception is the RT finger formed at the equator of the remnant. This RT finger maintains its radial shape and extents up to the level that penetrates and deforms the shocked shell behind the forward shock and shapes an ``ear'' into the overall SNR morphology. The formed equatorial ``ear'' is maintained in full growth for about 1000~yrs.  
 
After that period of time, the ``ear'' starts progressively to be swallowed by the main shell and the remnant approaches a spherical symmetry (Fig. \ref{fig:SNR_evol} g,h). The equatorial RT finger gets homogenised and dissipates within the surrounding gas and thus, it stars to be hardly distinctive. Finally, after 2800 yrs of evolution the lobe has been entirely engulfed within the remnant (Fig. \ref{fig:SNR_evol}i). At that moment the reverse shock has reached the center of the SNR while the forward shock has almost established a spherical shape. Thus, no information about the CSM interaction is carried anymore by the SNR.  


\subsection{``Ears'' formation mechanism }\label{subsec:ear_mech}

To further explain the formation and preservation mechanism of the extended equatorial RT-finger, i.e. the mechanism responsible for the  genesis and evolution of the SNR ``ears'', we illustrate in Fig.~\ref{fig:ear_formation}a the snapshot of the SNR evolution at $t_{\rm SNR}= 400$~yrs (i.e. same as Fig.~\ref{fig:SNR_evol}d) but zoomed in the region of the lobe.  In this plot the 2D density contours are accompanied by the velocity vectors of the  gas. In addition, in Fig.~\ref{fig:ear_formation}b, we present the 2D contours of the SNR shocked gas pressure. 

As it is  depicted in these plots, the initial rapid expansion of the SNR equatorial bulge and its propagation in the low density ISM causes a substantial drop of the post-shock pressure. As the pressure behind the lobe becomes lower than this of the neighboring gas, an angular pressure gradient is established that in turn triggers a tangential component at the shocked gas velocity towards the equator of the remnant. This tangential flow converges at the equatorial RT finger, supporting it to maintain its radial structure and grow further. The velocity of the gas that consists the RT finger is higher than this of the surrounding flow and as a result a Kelvin-Helmholtz instability grows at the outer part of the finger, forming an arrow-shaped tip. The compressed gas between the equatorial RT finger and the SNR's forward shock forces the latter to gain and retain its ``ear'' morphology for several hundreds up to a few thousand years after the SN explosion.  

 Finally, Fig. \ref{fig:ear_formation}c illustrates the spatial distribution of the three components that constitute the resulting SNR: the SN ejecta, the swept up wind material and the shocked ISM. Overall, as expected the SN ejecta dominates in the inner region of the SNR surrounded by the shocked wind material that has been accumulated into a thin shell. The outer layer of the SNR consists of the shocked ISM that is lying behind the remnant's forward shock. The three layer stratification is disturbed by the the RT instabilities where a partial mixing between the tree components occurs in the region of the contact discontinuity. In the area where the lobe has been shaped a different image emerges. Due to the equatorially confined circumstellar structure that was surrounding the explosion center, a large amount of CSM material that has been assembled into a thick dense region just behind the region of the lobe and substantial mixing occurs between the CSM and SN ejecta material. Finally, the extensive RT finger in the equator consists almost exclusively by CSM material surrounded by the shocked ISM.


\subsection{Models Comparison}\label{subsec:models_comp}

In this section, we evaluate differences in the morphology and properties of the formed ``ears'' in SNRs by varying the main physical variables involved in the model. Given that the final outcome is determined by the combination of the SN, CSM and ISM properties, the overall parameter space in our modeling is immense. For this reason, we eliminate our study by focusing on the SN explosion properties, the stellar wind properties and finally, these of the CSM bipolarity. In particular, we produced seven different models (in addition to  the model presented in Sect. \ref{subsec:SNR_hydro}; hereafter Model A) by varying: i) the SN ejecta mass and density power law index, ii) the stellar wind mass loss rate and iii) the bipolar CSM density and velocity distribution as defined by the constants $\alpha$, $\beta$ and $k$  (see Equations~\ref{eq:u_theta}~and~\ref{eq:rho_theta}). The parametric space we examined was centered on Model A, where the other seven models were produced by changing one of the aforementioned variables each time.

\begin{table*}
\centering
\caption {The eight studied SNR models interacting with bipolar circumstellar structure. The first column denotes the model's name and the second one the parameter that has been changed in respect to Model A. Columns (3)-(8) display the adopted parameters of each model where the changed parameters are denoted in bold. Finally, columns nine and ten tabulate the Figure number where the CSM and the SNR of each model are illustrated, respectively. In all models we considered an ISM density of $n_{ism}= 0.1~\rm cm^{-3}$ and temperature $T_{ism}=1000~\rm K$, while we let the CSM bubble to evolve for $\tau_w= 0.15$~Myr adopting a polar wind velocity of $u_{w,p}= 10~{\rm km~s^{-1}}$.}
\label{Table:CSM_comparison}
\begin{tabular}{ |c|c|c|c|c|c|c|c|c|c|  }
 \hline
 \multicolumn{10}{|c|}{Models Comparison Properties}\\
 \hline
 \hline
  (1)&(2)&(3)&(4)&(5)&(6)&(7)&(8)&(9)&(10)\\
 \hline
 \hline
 Model Name  & Parameter Changed  &  $M_{SN}~\rm (M_{\odot})$  & $n_{SN}~\rm (cm^{-3})$ & $\dot{M}_p~\rm (M_{\odot}~yr^{-1})$ & $k$ & $\alpha$ & $\beta$  &CSM & SNR   \\
\hline
 A& - & 1.38 & 7 & $2\times10^{-6}$ & 3 & 0.75 & 3 & Fig. \ref{fig:CSM_comparison}a & Fig. \ref{fig:SNR_comparison}a \\
 \hline
 B& Total SN ejecta mass & {\bf 7} & 7 & $2\times10^{-6}$ & 3 & 0.75  & 0.8 & Fig. \ref{fig:CSM_comparison}a & Fig. \ref{fig:SNR_comparison}b \\
 \hline
 C& SN density power law index & {\bf 7} & {\bf 11} & $2\times10^{-6}$ & 3 & 0.75 & 0.8 & Fig. \ref{fig:CSM_comparison}a & Fig. \ref{fig:SNR_comparison}c \\
 \hline
 D& Stellar wind mass loss rate & 1.38 & 7 & $ {\bf 1\times10^{-5}} $ & 3 & 0.75 & 0.8 &  Fig. \ref{fig:CSM_comparison}b & Fig. \ref{fig:SNR_comparison}d \\
 \hline
 E& CSM equatorial confinement  & 1.38 & 7 & $1.5\times10^{-6}$ & {\bf 1} & 0.75 & 0.8 & Fig. \ref{fig:CSM_comparison}c & Fig. \ref{fig:SNR_comparison}e \\
 \hline
 F& CSM equatorial confinement & 1.38 & 7 & $2.3\times10^{-6}$  & {\bf 6} & 0.75 & 0.8 & Fig. \ref{fig:CSM_comparison}d & Fig. \ref{fig:SNR_comparison}f \\
 \hline
 G& CSM polar to equatorial ratios  & 1.38 & 7 & $2.4\times10^{-6}$ & 3 & {\bf 0.65} & {\bf 0.7} &Fig. \ref{fig:CSM_comparison}e & Fig. \ref{fig:SNR_comparison}g \\
 \hline
 H&CSM polar to equatorial ratios  & 1.38 & 7 & $1.4\times10^{-6}$ & 3 & {\bf 0.85} & {\bf 0.9} &Fig. \ref{fig:CSM_comparison}f & Fig. \ref{fig:SNR_comparison}h \\
 \hline
\end{tabular}
\end{table*}

\begin{figure*}
\includegraphics[clip=true,width=\textwidth,angle=0]{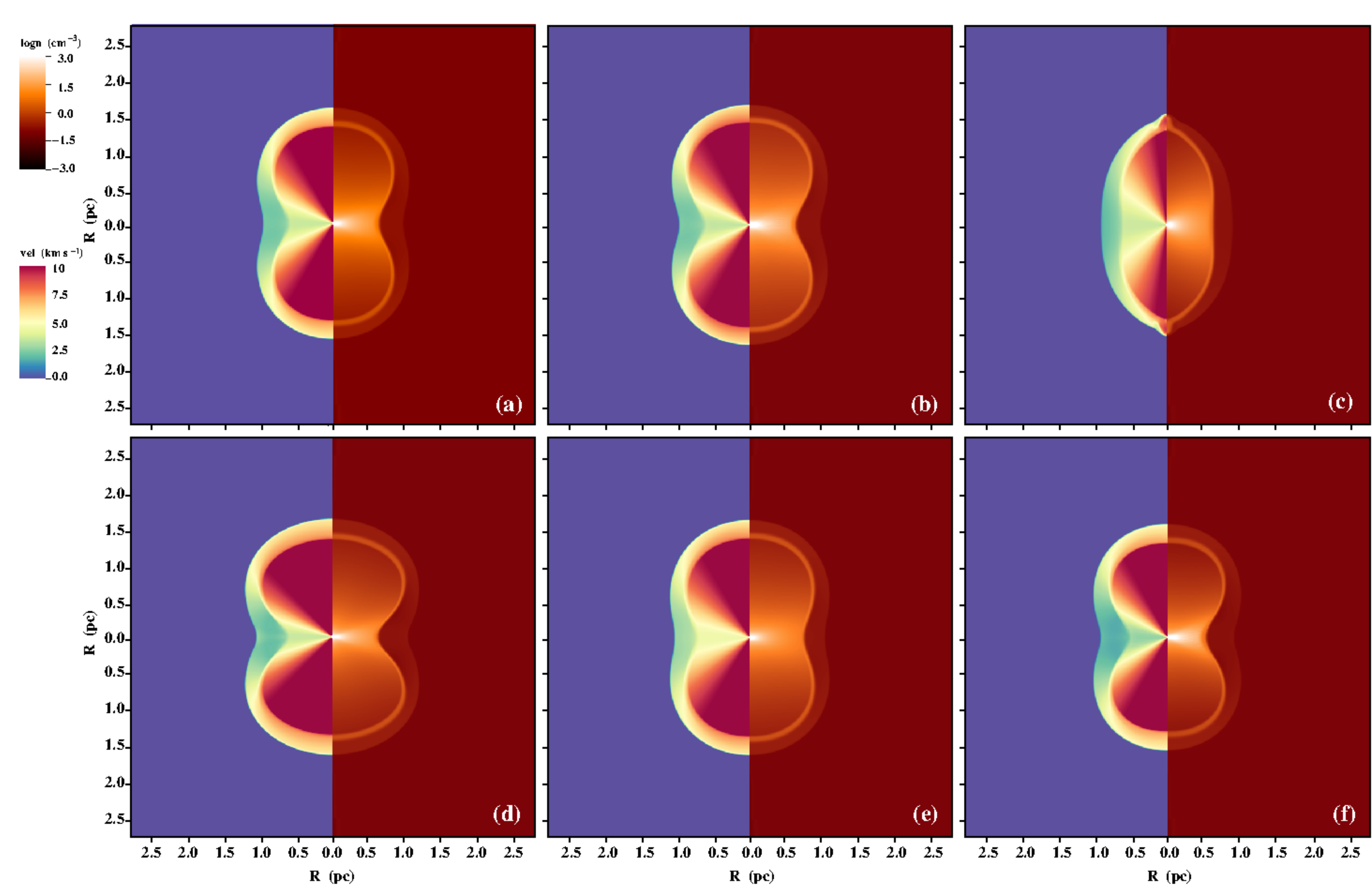}
\caption {The 2D density and velocity contours of the CSM at the moment of the SN explosion. (a): The CSM that corresponds to Models A, B and C and (b), (c), (d), (e), (f): the resulted CSM of  Models D, E, F, G and H, respectively (see Table \ref{Table:CSM_comparison} and text for details).}
  \label{fig:CSM_comparison}
\end{figure*}

\begin{figure*}
\includegraphics[clip=true,width=126mm,angle=0]{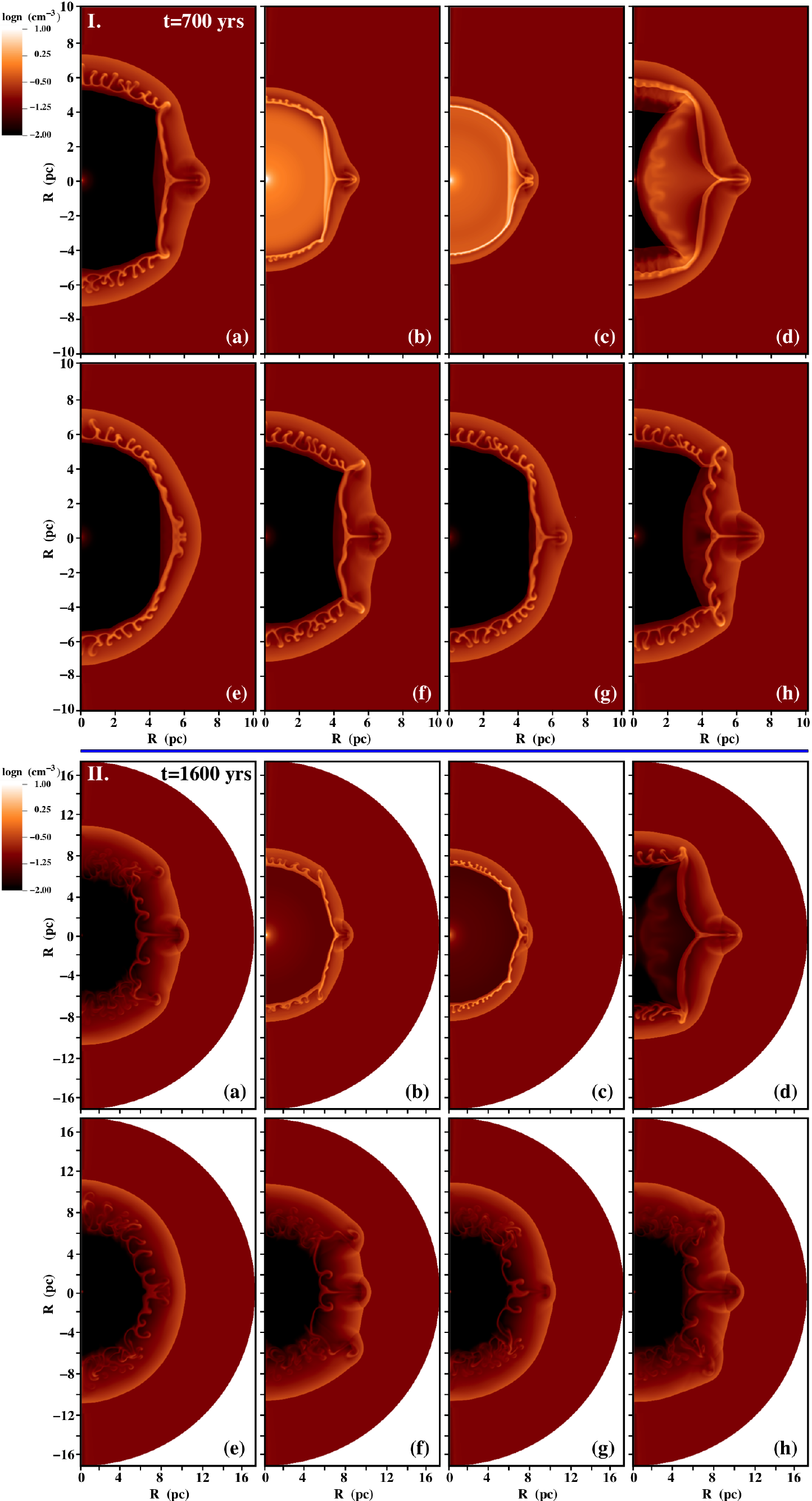}
\caption {The 2D density contours of the SNR for Models A-H. The upper sequence of plots (I) correspond to the SNR age of t= 700 yrs, while the lower one for t= 1600 yrs. See Table \ref{Table:CSM_comparison} and text for details.  }
  \label{fig:SNR_comparison}
\end{figure*}

The properties adopted for the eight models are summarised in Table \ref{Table:CSM_comparison}. Figure \ref{fig:CSM_comparison} illustrates the CSM density and velocity distribution at the moment of the SN explosion,  while Figure~\ref{fig:SNR_comparison} presents the resulting SNR, for each model, at two different snapshots: $t_{\rm SNR}$= 700 yrs (Fig. \ref{fig:SNR_comparison}I) and $t_{\rm SNR}$=  1600 yrs (Fig. \ref{fig:SNR_comparison}II). The main conclusions resulted by our hydrodynamic simulations and the subsequent comparison among the models are the following: 
\\
\\
\underline{{\it Varying the SN explosion properties (Models B \& C) }}: In this set of models we kept the bipolar CSM structure the same as Model~A (Fig. \ref{fig:CSM_comparison}a) and we changed the SN explosion properties.  We first increased the SN ejecta mas from 1.38 $\rm M_{\odot}$ to 7 $\rm M_{\odot}$ (Model B) and subsequently, we run an additional model with a higher power law index of the SN density profile ($n_{SN}= 11$;  Model C). The adopted parameters of Model B could potentially correspond to a Type Ib/c SN, while those of Model C to a Type II event, as the latter are characterised by steeper outer density profiles \citep{Chevalier1982b}.

The SNR resulted by {\bf Model B} is depicted in Fig. \ref{fig:SNR_comparison}b,I-II. Due to the higher ejecta mass -as compared to Model A- the SNR evolves slower while the shocked ejecta shell and the freely expanding ejecta region reveal  higher densities. Within the timescales of our simulations, no substantial deceleration occurs in the SNR  and the reverse shock remains relatively close to the contact discontinuity.  As a result the growth of the RT instabilities remains limited. The equatorial RT finger displays a similar behaviour with Model A, being radially extended and forcing the remnant to shape a local ``ear''.  After its initial growth (Fig. \ref{fig:SNR_comparison}b,I), the lobe starts progressively to be merged within the shocked ambient medium shell (Fig. \ref{fig:SNR_comparison}b,II). Nevertheless, contrary to Model A, in Model B the local lobe dissipates when the SNR is still in the free expansion phase where the reverse shock is still active. In conclusion, for given CSM conditions, by changing the SN ejecta mass, no essential differences are expected to the ear's morphology and properties. However, as the SN mass increases the whole process of the ear formation, growth and dissipation occurs in prior stages of the SNR evolution. 

In {\bf Model C} we additionally increased the power law spectral index of the SN density profile ($n_{SN}=$ 11). As it is clearly illustrated  in Fig. \ref{fig:SNR_comparison}c,I-II the formed lobe is smaller in size compared to Models~A and B while after 1600 yrs is hardly distinctive.  Such a result is expected as by increasing the SN density power law index, the dependence of the SNR expansion velocity on the CSM density becomes weaker \citep{Blondin1996, Chevalier82}. Thus, the CSM density enhancement at the equator has moderate effects on the evolution of the subsequent SNR and as a result the phenomenon of the ``ears'' formation is mitigated.
\\
\\
\underline{{\it Varying the stellar wind properties (Model D)}}: In Model D we kept the same SN properties and the constants $\alpha, \beta, k$ that determine the CSM bipolarity as Model A but we increased the stellar wind mass loss rate by five times ($\dot{M}_p= 10^{-5}~\rm M_{\odot}~yr^{-1}$). As expected, the resulted CSM (Fig. \ref{fig:CSM_comparison}b) retains the same structure but it is denser and slightly more extended than the CSM of Model~A. 

The resulted SNR of {\bf Model D} is illustrated in Figure \ref{fig:SNR_comparison}d,I-II. Comparing to Model A, there are two main distinctive differences. The first one is the formation of a fast moving reflected shock, triggered by the collision of the SNR with the density walls of the wind bubble, that rapidly approaches the center of the remnant. The morphology of the reflected shock follows this of the wind bubble, being more evolved at the SNR's equator where a large amount of shocked CSM material has been accumulated. The second difference compared to Model~A  regards the ``ear'' properties.   The higher density contrast between the circumstellar structure  and the surrounding ISM, results to a stronger shock breakout of the SNR.  The high post shock pressure of the swept up CSM at the equatorial plane of the remnant is higher and  consequently, a more profound and extended lobe is formed at the resulting SNR. The shaped ``ear'' preserves its structure for a much larger time interval, being survived till the end of our hydrodynamic simulations (i.e. $t_{SNR}$=~2800~yrs). 
\\
\\
\underline{{\it Varying the properties of the CSM bipolarity (Models E-H)}}: In the sequence of Models E-H, we study the dependence of the resulting SNR properties on the level of bipolarity of the surrounding CSM. For this purpose we kept the same SN properties as Model~A. In addition, we normalised the stellar wind mass loss rate of each model in order to maintain the same total mass of the stellar wind enclosed in the CSM bubble as this of Model A. 

At Models E and F, the polar to equatorial CSM  density and velocity ratios are the same with Model A (i.e. we kept the same $\alpha, \beta$) but we changed the confinement level of the equatorial disk by adopting $k=$ 1 and 6, respectively. 

Due to the low  equatorial wind confinement of {\bf Model E}, the resulted wind bubble deviates from the bipolar morphology and displays a rather elliptical shape where the radius, the  density and the  velocity of the CSM change smoothly from the poles to the equator (Fig. \ref{fig:CSM_comparison}c). The interaction of the subsequent SNR with such a circumstellar structure is not able to form the local equatorial lobes. Instead, the remnant presents a small bulge on its equator (Fig. \ref{fig:SNR_comparison}e,I) that after 1600 yrs it has been completely disappeared (Fig. \ref{fig:SNR_comparison}e,II).

By contrast, the high value of $k$ imported in {\bf Model F} results on a bipolar CSM structure characterised by a narrow equatorial waist of dense, slow moving material (Fig. \ref{fig:CSM_comparison}d). Comparing the resulted SNRs of Models A and F there is no any noticable difference  regarding the size, the geometry and the life duration of the formed ``ears''. Nevertheless, a novel  feature that emerges from the SNR of Model F is the formation of two additional extended RT fingers that arise antisymmetrically to the equatorial one at about $\theta= 60^o$ and $120^o$ (Fig. \ref{fig:SNR_comparison}f,I). Similar to their equatorial counterpart, the two RT fingers penetrate the shell of the shocked gas,  deform the SNR forward shock and finally, shape two additional lobes on the overall morphology of the remnant (Fig. \ref{fig:SNR_comparison}f,II). 

The formation of these two RT fingers -that are also present to the rest models but to a lower extend- is attributed to the obliquity of the reverse shock to the freely expanding SN ejecta at the point where the CSM bubble bends to form the equatorial waist. As it has been shown also by \citet{Blondin1996} the non-frontal collision  of the freely expanding ejecta with the reverse shock provokes strong vorticity on the post-shock flow that in turn supports the growth of an extended RT instability. 

The final models of the studied set are the Models G and H.  
 In these models the parameter that has been changed - compared always to Model A- is the ratio of the CSM density and  velocity  from the poles to the equator as defined by the constants $\alpha$ and $\beta$.

Figure \ref{fig:CSM_comparison}e  shows the CSM of {\bf Model G}. Due to the low $\alpha$ and $\beta$ values imported in this model, the wind bubble reveals a moderate bipolar shape with a equatorial radius being $\sim 20 \%$ smaller than the polar one. Even if for such a circumstellar structure are met the required criteria for the ``ear'' formation on the resulting SNR, the formed lobe is hardly distinguished after 700 yrs of evolution (Fig. \ref{fig:SNR_comparison}g,I) and it is completely engulfed within the shocked ambient medium shell after 1600 yrs (Fig. \ref{fig:SNR_comparison}g,II). 

Finally, Fig. \ref{fig:CSM_comparison}f illustrates the CSM of {\bf Model H}, characterised by a typical elongated hourglass with a narrow, dense waist. The interaction of the SN ejecta with this circumstellar structure shapes a sizable ``ear'' much larger in both length and width  compared to Model A (Fig. \ref{fig:SNR_comparison}h,I). Furthermore, the two additional lobes met in Model F are also present in Model H (Fig. \ref{fig:SNR_comparison}h,II). However, the position of the two additional lobes is shifted more towards the poles of the remnant (placed at $\theta= 50^o~{\rm and}~130^o$), while they remain smaller than the equatorial lobe.


\section{Discussion}\label{sec:Discuss}

\subsection{Comparing the models' results to the relevant observations}
\label{subsec:ModelVsObs}

We have argued that the ``ears'' observed in several SNRs were formed by the interaction of the SN ejecta with a bipolar CSM that occurred during the early phases of the SNR evolution. Through the conduction of the hydrodynamic simulations was intended to present the physical mechanism lying behind the ``ears'' formation and demonstrate the dependence of the final outcome on the SN/CSM properties. Thus, in our modeling we did not aim to model any specific SNR -a process that requires a thorough modeling and detailed fine-tuning of the imported parameters- whereas the chosen parametric space imported in our models was limited into a specific range  that definitely can not cover all cases of SNRs with ``ears''.  Nevertheless, the results extracted by  the set of models presented in this work reveal a sequence of intriguing similarities with the properties of the ``ears'' observed in nearby SNRs. 

The first similarity regards the size of the formed ``ears''. In our models the radius of the remnant is about  $10\%$ to $65 \%$ larger in the region of the formed lobe than its overall radius (with Model E revealing the smallest  ear radius and Model H the largest one). In the known sample of SNRs that have been observed to host ``ears'' in their morphology (see Fig. \ref{fig:SNRs_ears} and \citet{Bear2017}; \citet{Tsebrenko2015b}), the nebula's radius at the region of the lobe lengths up by $ \sim 30 - 50 \%$ as compared to the radius of the main shell i.e. being well within the range of our extracted results.  Same applies for the timescales of the ``ears'' lifespan resulted by our modeling. We have shown that the ``ears'' start to be formed in about 100 years after the SN explosion (Fig. \ref{fig:SNR_evol}c) and in many cases they survived till the end of our hydrodynamic simulation i.e. $t_{SNR}= 2800$ yrs. This time range covers most of the SNRs ages observed to reveal ``ears'' in their morphology. Exception is the SNR S147 whose kinematical age is estimated to be about an order of magnitude larger \citep{Kramer2003}. That means that either S147 encountered a much larger and/or denser ciscumstellar structure than these presented in this work or  another physical mechanism was responsible for the formation of its ``ears'' (see Sect. \ref{subsec:jetVsbipolar comparison}).

Our modeling extracted a range of ``ears'' morphologies depending on the selected CSM/SN properties and/or the evolutionary phase of the SNR. In particular, it has been shown that when the SNR is still young ($80 {\rm yrs} < t_{SNR}$ < 400 {\rm yrs}; Fig. \ref{fig:SNR_evol}c) or the CSM is characterized by a modest density gradient from the poles to the equator (Models E and G; Fig. \ref{fig:SNR_comparison}e,g) the resulted ``ear'' appears as a bulge of an almost triangular shape that progressively starts to protrude as we are moving toward the equator of the remnant. By contrast, for the cases where the SNR is older than $\sim 400$~yrs (Fig. \ref{fig:SNR_evol}d,e,f) and for CSM structures that display a well confined, narrow and dense equatorial waste (e.g. Models D and H; Fig. \ref{fig:SNR_comparison}d,h), the formed ``ears'' reveal the morphology of a localised lobe that inflates out of the SNR's main shell. The  former morphology (triangular shape) is more similar to this observed in the very young ($t\sim 100$ yrs) SNR G1.9+0.3, while the latter (local lobe) is met in most SNRs with ``ears'' such as Kepler's SNR, G309.2-06 and S147 (see Fig. \ref{fig:SNRs_ears}).

Another point extracted by our simulations is that during the early phases of the ``ear'' formation ($t_{SNR} < 700$~yrs) the remnant' s forward shock in the region of the lobe is faster than the overall expansion velocity of the SNR. Such a result is aligned to the bright X-ray synchrotron emission observed in front of the ``ears'' at the young SNRs of Kepler's \citep{Vink2008} and G1.9+0.3 \citep{Borkowski2017}, which indicates that the remnant reveal high expansion velocities in these regions. Regarding Kepler's SNR, it arises another interesting similarity to the results of our modeling. X-ray and Infrared observations of the remnant \citep[][respectively]{Burkey2013,Williams2012} have showed that the central regions of the SNR are occupied by dense, shocked CSM material, which is lying on a strip roughly along the site that connects the two ``ears''. As we have shown in Figure \ref{fig:ear_formation}c, our models predict the accumulation of a large amount of CSM at the equator of the remnant i.e. where the two ``ears'' are sculptured, in agreement with what is observed on Kepler's SNR. 

Finally, an interesting feature resulted by our modeling is that under specific CSM conditions (Models F and H), apart from the equatorial lobe, two additional lobes are formed in the final morphology of the remnant. Intriguingly, such a three-lobe structure has been observed in the South-West region of the LMC SNR DEM~34A \citep{Meaburn1987}, which  was attributed by the author to local inhomogeneities within the parent sheet  or alternatively to the collision of the blast wave with a helical annulus. Our model offers an alternative explanation for the formation of these three-lobe structure observed in DEM~34A. Same applies for the case of N63A  that reveals a multi-lobe morphology consisting of ``crescent''-shaped lobes of several sizes \citep{Warren2003}. Of course as mentioned above, detailed modeling -most likely involving extra ingredients than the  axis-symmetric model presented here- is required in order to reproduce the ``exotic'' morphologies of the aforementioned SNRs.


\subsection{Bipolar CSM structures around supernova progenitors}
\label{subsec:bipolar_origin}

\begin{figure}  
\includegraphics[clip=true,width=\columnwidth,angle=0]{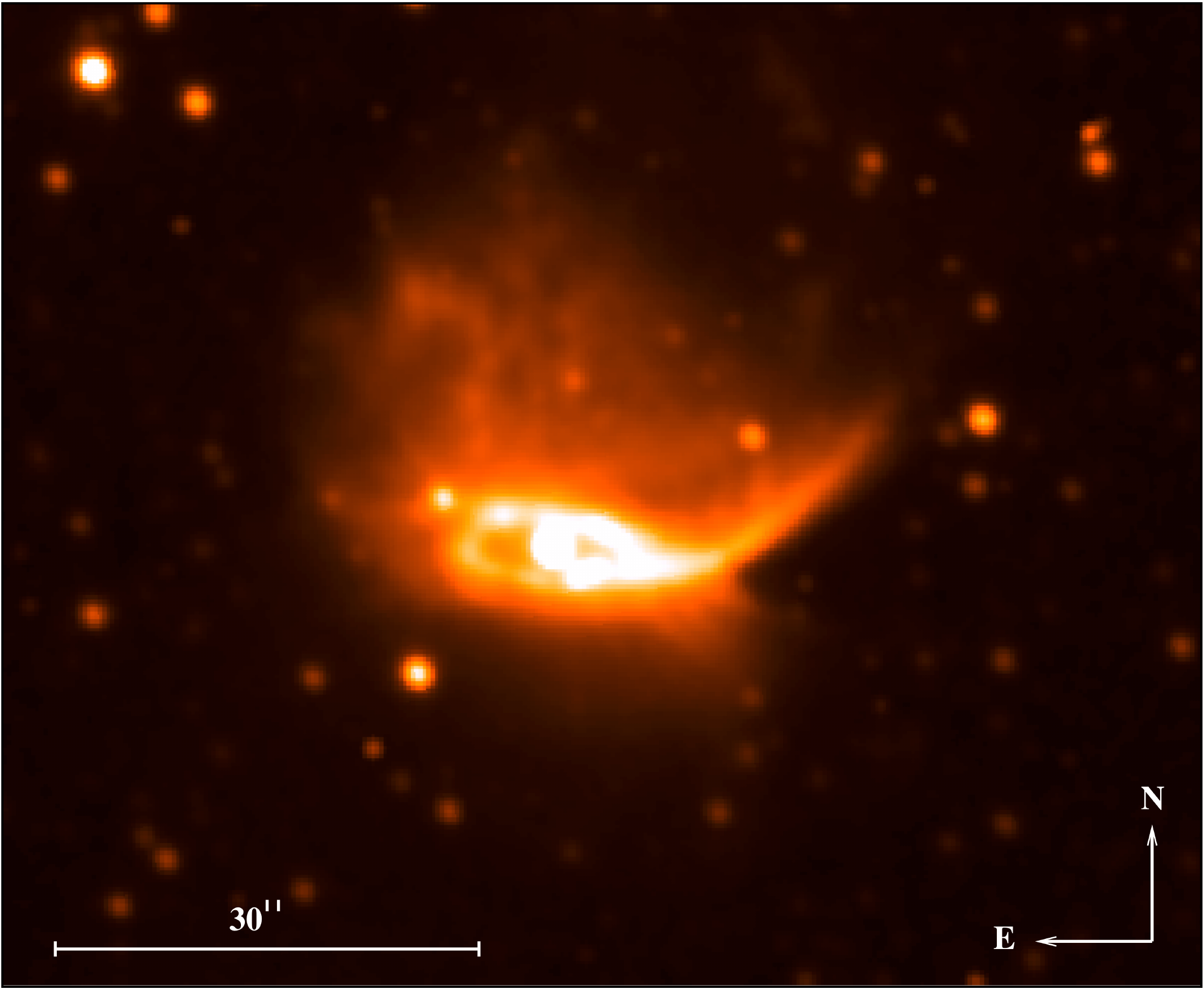} 
\caption {The ${\rm H{\alpha}}$ image of the bipolar PN Henize 2-428 as observed by the 2.3m Aristarchos telescope at Helmos Observatory (Greece) on Novembber 12, 2015. The detector was a 2048$\times$2048, 13.5 $\mu$m pixels
CCD, with a field of view of 5.5$\times$5.5 \,arcmin$^2$ (0.32 \,arcsec \,pixel$^{-1}$ in binning 2$\times$2), while the exposure time was 1800 s.}
  \label{fig:Henize}
\end{figure}

The fundamental assumption of our model is that the SN explosion center is surrounded by a dense circumstellar bubble characterised by a bipolar morphology and density enhancement at its equatorial plane. The existence of such CSM structures is predicted by the stellar evolution theory and has been confirmed by numerous observations. 

Indeed, a significant fraction of evolved stars -throughout the Hertzsprung Russel diagram- have been found to be engulfed by nebulae that possess bipolar symmetry. These circumstellar structures are characterised by dense equatorial disks, rings or tori that form a narrow, dense waist around the central star and two extended polar cups that are lying perpendicular to the equatorial plane (e.g. Fig. \ref{fig:Henize}). The physical processes responsible for the formation of the bipolar nebulae around evolved stars are still debatable. Stellar rotation \citep[e.g.][]{Heger2000, Bjorkman1993}, close binary interactions  \citep[e.g.][]{Politano2011, Huggins2009, Mastrodemos1999}, eruptive mass loss \citep[e.g.][]{Smith2014} and magnetic fields  \citep[e.g.][]{Townsend2005,Matt2004, GarciaSegura1999} have been suggested as the most eminent mechanisms to explain the stellar outflows confinement towards the equator and the formation of bipolar circumstellar structures.

Regarding the high stellar mass regime, the types of stars that have most frequently been found to be surrounded by bipolar CSM are the Luminous Blue Variables  \citep[LBV, e.g. $\eta$ Carinae;][]{Smith2002} and the Blue Supergiants \citep[BSG, e.g. SBW1, SBW2;][]{Smith2007}. Indications of bipolar and/or disk like CSM morphologies are also met in colder massive stars such as Yellow Supergiants (YSG) and evolved Red Supergiants (RSG) like the IRC +10420 \citep{Tiffany2010}. Given that all these types of stars are potential Core Collapse SN progenitors, it is expected that a percentage of SNe occur within bipolar circumstellar structures and subsequently interact with them. The most characteristic example of such a SN explosion is SN 1987A  \citep{West1987}, a Type II SN with a BSG progenitor \citep{Hillebrandt1987}. The CSM around this SN shows a dense equatorial ring surrounded by two polar cups \citep{Plait1995,Crotts1991}. Intriguingly, the CSM of SN 1987A is almost identical with that found around the B1.5Ia supergiant Sher~25 \citep{Brandner1997} and the Galactic LBV HD168625 \citep{Smith2007b}. Other cases of core collapse SNe that show evidence of interaction with dense bipolar structures are the Type IIn SNe SN~2010jl \citep{Katsuda2016} and SN~2009ip \citep[]{Mauerhan2014, Reilly2017}, for which it has been suggested that they originated from an LBV progenitor. SN~1998S is an additional observational evidence of a Type IIn CC SN with a bipolar CSM coming most likely from an evolved red or yellow supergiant star \citep{Chugai2001}.

As far as the low mass stars is concerned, signs of aspherical CSM that display bipolar symmetry are mostly found in PNe \citep[e.g. Hb12, MyCn 18, Mz 3,][respectivelly]{Clark2014,Conor2000,Clyne2015}. Very recently, through the ALMA array, a sample of stars in the Asymptotic Giant Branch (AGB) phase  was observed in high resolution and it was found that their winds sculpture non-spherical circumstellar structures which share common properties with these of PNe i.e. bipolar symmetry hosting a density enhanced equatorial waist \citep{Decin2020S}. AGB stars have been proposed as possible candidates for the donor stars of the SNe Ia progenitor systems, especially for the cases where these cosmic explosions reveal ${\rm H{\alpha}}$  narrow emission lines during their early spectra \citep[e.g. SN 2002ic;][]{Hamuy2003}. In addition, growing theoretical and observational evidence appear in the literature showing that a fraction of SNe Ia occur in the center of PNe shaped by their progenitor systems \citep{Chiotellis2020,Cikota2017,Tsebrenko2015c,Santander2015}. Thus, apart from their core collapse counterparts, SNe Ia are also likely to occur within and subsequently interact with a bipolar CSM. 

Finally, indirect evidence of SN interaction with bipolar circumstellar structures are shown through the observables we receive from a number of nearby SNRs. It has been suggested that their non spherically symmetric morphology, kinematics and emission properties can best explained if the progenitor stellar system had mass outflows confined more to the equatorial plane than along the poles \citep{Burkey2013,Gaensler1998, Igumenshchev1992}. 

All the aforementioned evidence point towards the plausibility of our SNR ``ear'' formation model in nature, as they advocate for the fact that a non negligible percentage of supernovae (both core collapse and Type Ia) are surrounded by equatorially confined bipolar CSM. The second important assumption of our model is that the formed circumstellar structures is denser than the ambient medium. This condition is required for the SNR shock breakout, which in the framework of our model is the central engine for the formation of the two antisymmetric ``ears''. Hence, our model is more aligned to SN progenitors, which during their final phases of evolution where suffering by intense and slow mass outflows.  In conclusion, if indeed the two lobes in a SNR have been formed under the physical process suggested in this model, the most plausible progenitor of the remnant is a LBV or a YSG/RSG star for the case of core collapse SNe and a symbiotic binary system or a PN for the SNe Ia. Nevertheless, other important factors such as stellar duplicity and/or the involvement of  magnetic fields can substantially alter the evolution and mass outflows properties of the progenitor system and thus, broad the  spectra of stellar candidates suggested above.

\subsection{Comparison with the jet and Blondin et al. 's model}\label{subsec:jetVsbipolar comparison}

We have presented evidence that the antisymmetric ``ears'' observed in a number of SNRs, could potentially be sculptured through the SNR shock breakout from a bipolar CSM that was surrounding the explosion center. This scenario is placed as a third alternative to the other two outstanding ``ears'' formation models suggested in the literature: a) the jet model, which suggests that the ``ears'' are shaped by the interaction of the SNR with two jets launched by the progenitor system \citep[e.g.][]{Tsebrenko2013} and b) the model of \citet{Blondin1996} in which the ``ears'' are inflated during the evolution of the SNR within an axisymmetric and equatorially enhanced circumstellar structure. 

These three ear formation models seem to be complementary rather than mutually exclusive, as they rely on different physical mechanisms that are independent from each other and thus, they are applicable in different systems. For instance, in a number of SNRs that reveal ``ears'' on their morphology, one or two jets have been observed lying within the direction of the two opposite lobes \citep[e.g. W50;][]{Fabrika2004}. Such an evidence points directly towards the jet formation model. On the other hand, evolved Type Ia SNRs are difficult to be explained under the jet model as in the vast majority of SNe Ia explosions the white dwarf progenitor is completely destroyed \citep{Livio2018} so no stellar system is remained to trigger a jet. 

Luckily, there are two main differences among the three models that to a large extent can make the physical origin of the formed ``ear'' on SNRs distinctive. These two differences are: a) the orientation of the formed protrusions in respect to the host SNR  and b) the timescales in which the ear formation process occurs regarding the  SNR evolutionary phase.  
\\
\underline{\it Ears' orientation}: In the model of \citet{Blondin1996}, the two protrusions emerge along the polar axis of the SNR where the CSM displays the lowest density. Similarly, given that the jets are expected to be launched perpendicularly to the equator of the stellar system, the two ears are more likely to be inflated close to the poles of the remnant. In this model the two antisymmetric lobes are shaped in the equator of the SNR where the CSM is confined into a dense and narrow waist. Consequently, if the two lobes of a SNR have been formed by the shock breakout from a bipolar CSM -i.e. as suggested by this model- a large amount of shocked CSM material is expected to be concentrated within a strip between the two ears (as observed in the case of Kepler's SNR; see Sect. \ref{subsec:ModelVsObs}).  By contrast, in \citet{Blondin1996} and jet model the region of the dense, shocked CSM should lay roughly perpendicular to the line that connects the two lobes. 
\\
\underline{\it Involved timescales}: The model of \citet{Blondin1996} predicts that the two protrusions are formed when the SNR is {\it within} the axisymmetric circumstellar bubble. Thus, it is expected that two opposed ``ears'' are shaped when the remnant is still in a  very early evolutionary state. Such a model is sufficient to explain young SNRs such as the 41.9+58 in M82 \citep{Bartel1987}, but it seems difficult to be applied on mature SNRs that have been evolved beyond the adiabatic phase  (e.g. S 147, G309.2-0.6). On the other hand, the model presented in this work advocates that the two opposite lobes are sculptured when the remnant has utterly penetrated the circumstellar structure and starts to propagate into the ambient medium. The time interval in which the two ``ears'' are present ranges from hundreds to thousands of years after the SN explosion, while for given SN/CSM properties can last well beyond the Sedov phase. Hence, the two axisymmetric CSM models -\citet{Blondin1996} and the present one- cover complementary different evolutionary stages of SNRs. Finally, regarding the jet model it is difficult to predict the phase in which the ``ears'' are formed, given that  two jets can be launched from the progenitor system at any time of the SNR evolution  \citep{Bear2017, Grichener2017}. Nevertheless, as mentioned above, for the case of Type Ia SNe the jets  can not be launched after the SN explosion, something that places strict spatial and temporal limits on the ``ear'' formation process.  In particular, the SNe Ia jet models \citep{Tsebrenko2013} suggest that the ``ears'' were formed through the collision of a) a `jet-carrying' SN ejecta with a spherical (pre-)PN shell or b) a spherical SN ejecta with an `ear-shaped' PN shell, depending on whether the jets are triggered during or prior  the explosion, respectively. Thus, according to these models the spatial scales for the ``ear'' formation should be comparable to the size of the surrounding (pre-)PN. Given that the majority of PNe reveal a radius of  $R_{\rm PN} \sim $ 0.05 – 1 pc  \citep{Frew2016}, a similar spatial-scale is imposed for the SNe Ia jet model. 


\section{Summary}\label{sec:Summary}

The results and the main conclusions of this work are summarized as follows.

\begin{enumerate}

\item We modeled the interaction of a SNR with a surrounding bipolar, equatorially confined circumstellar structure. We have shown that such an interaction can be responsible for the formation of two opposite lobes (``ears'') frequently met in the morphology of several SNRs.

\item According to our model, the ``ears'' are formed by the SNR's forward shock breakout from  the bipolar CSM.  Due to the geometry of the SNR/CSM system,  the post shock flow converge at the equator of the remnant, something that supports the growth of an extended RT finger. This RT finger presses and deforms the SNR's forward shock shaping a local lobe in the equator of the remnant. This lobe survives from hundreds to thousands of years after the SNR's shock breakout from the CSM.

\item We ran a set of models evaluating differences in the morphology and the properties of the formed SNR ``ears'' by varying the CSM and SN properties. We found that both the size and the lifespan of the formed ``ears'' increases with a) the density contrast between the circumstellar bubble and the ambient ISM, b) the level of confinement of the circumstellar material at the equator of the system, and c) the ratio of the CSM polar density  and flow velocity over the equatorial ones. Regarding the SN properties, we found that by increasing the ejecta mass, the ``ears'' are present on earlier evolutionary stages of the SNR, while the protrusion is less pronounced for SNe characterised by sharp declining ejecta density profiles. 

\item The results extracted from the grid of our hydrodynamic simulations reveal a number of similarities to the relevant observables regarding the size, the lifespan and the  kinematics of the ``ears'' in SNRs. In addition, our model predicts the accumulation of a large amount of shocked CSM in the region between the two antisymmetric lobes something that has been observed in the case of Kepler's SNR. Finally, under specific CSM conditions our modeling extracted SNRs that possess six lobes in their morphology. Such multi-lobes features has been observed in a number of SNRs such as the DEM~34A.

\item We presented theoretical and observational evidences of supernova explosions occurring within a bipolar CSM adopted in our model. This fact  enhances the plausibility of our model in nature. We discussed the most possible  progenitors of the SNRs that reveal two opposite lobes, under the framework of our model, suggesting to be the YSGs/RSGs or LBVs for the core collapse SNe and the symbiotic binaries or the PNe for the thermonuclear SNe. 

\item We compared our model with the other `ear-formation models' suggested in the literature. We showed that these models can be distinctive as they predict different orientation of the formed ``ears'' in respect the host SNR and different time scales in which the ear formation process occurs. 

\end{enumerate}

\section*{Acknowledgments}

This research is co-financed by Greece and the European Union (European Social Fund-ESF) through the Operational Programme `Human Resources Development, Education and Lifelong Learning 2014-2020' in the context of the project `On the interaction of Type Ia Supernovae with Planetary Nebulae' (MIS~5049922). A.C. is grateful to Noam Soker for many helpful discussions on the topic of jets interaction with supernova remnants.  In addition,  we thank Rony Keppens for providing us with the AMRVAC code. A.C. acknowledge the support of this work by the project `PROTEAS II' (MIS 5002515),which is implemented under the Action `Reinforcement of the Research and Innovation Infrastructure', funded by the Operational Programme `Competitiveness, Entrepreneur- ship and Innovation` (NSRF 2014-2020)
and co-financed by Greece and the European Union (European Regional
Development Fund). The Aristarchos telescope is operated on Helmos Observatory
by the IAASARS of the National Observatory of Athens.

\section*{DATA AVAILABILITY}

The data underlying this article will be shared on reasonable request to the corresponding author.

\bibliographystyle{mnras}
\bibliography{SNRs}

\begin{thebibliography}{}
\makeatletter
\relax
\def\mn@urlcharsother{\let\do\@makeother \do\$\do\&\do\#\do\^\do\_\do\%\do\~}
\def\mn@doi{\begingroup\mn@urlcharsother \@ifnextchar [ {\mn@doi@}
  {\mn@doi@[]}}
\def\mn@doi@[#1]#2{\def\@tempa{#1}\ifx\@tempa\@empty \href
  {http://dx.doi.org/#2} {doi:#2}\else \href {http://dx.doi.org/#2} {#1}\fi
  \endgroup}
\def\mn@eprint#1#2{\mn@eprint@#1:#2::\@nil}
\def\mn@eprint@arXiv#1{\href {http://arxiv.org/abs/#1} {{\tt arXiv:#1}}}
\def\mn@eprint@dblp#1{\href {http://dblp.uni-trier.de/rec/bibtex/#1.xml}
  {dblp:#1}}
\def\mn@eprint@#1:#2:#3:#4\@nil{\def\@tempa {#1}\def\@tempb {#2}\def\@tempc
  {#3}\ifx \@tempc \@empty \let \@tempc \@tempb \let \@tempb \@tempa \fi \ifx
  \@tempb \@empty \def\@tempb {arXiv}\fi \@ifundefined
  {mn@eprint@\@tempb}{\@tempb:\@tempc}{\expandafter \expandafter \csname
  mn@eprint@\@tempb\endcsname \expandafter{\@tempc}}}

\bibitem[\protect\citeauthoryear{{Badenes}, {Borkowski}, {Hughes}, {Hwang}  \&
  {Bravo}}{{Badenes} et~al.}{2006}]{Badenes2006}
{Badenes} C.,  {Borkowski} K.~J.,  {Hughes} J.~P.,  {Hwang} U.,   {Bravo} E.,
  2006, \mn@doi [\apj] {10.1086/504399}, \href
  {https://ui.adsabs.harvard.edu/abs/2006ApJ...645.1373B} {645, 1373}

\bibitem[\protect\citeauthoryear{{Bartel} et~al.,}{{Bartel}
  et~al.}{1987}]{Bartel1987}
{Bartel} N.,  et~al., 1987, \mn@doi [\apj] {10.1086/165847}, \href
  {https://ui.adsabs.harvard.edu/abs/1987ApJ...323..505B} {323, 505}

\bibitem[\protect\citeauthoryear{{Bear}, {Grichener}  \& {Soker}}{{Bear}
  et~al.}{2017}]{Bear2017}
{Bear} E.,  {Grichener} A.,   {Soker} N.,  2017, \mn@doi [\mnras]
  {10.1093/mnras/stx2125}, \href
  {https://ui.adsabs.harvard.edu/abs/2017MNRAS.472.1770B} {472, 1770}

\bibitem[\protect\citeauthoryear{{Bjorkman} \& {Cassinelli}}{{Bjorkman} \&
  {Cassinelli}}{1993}]{Bjorkman1993}
{Bjorkman} J.~E.,  {Cassinelli} J.~P.,  1993, \mn@doi [\apj] {10.1086/172676},
  \href {https://ui.adsabs.harvard.edu/abs/1993ApJ...409..429B} {409, 429}

\bibitem[\protect\citeauthoryear{{Blondin}, {Lundqvist}  \&
  {Chevalier}}{{Blondin} et~al.}{1996}]{Blondin1996}
{Blondin} J.~M.,  {Lundqvist} P.,   {Chevalier} R.~A.,  1996, \mn@doi [\apj]
  {10.1086/178060}, \href
  {https://ui.adsabs.harvard.edu/abs/1996ApJ...472..257B} {472, 257}

\bibitem[\protect\citeauthoryear{{Borkowski}, {Reynolds}, {Hwang}, {Green},
  {Petre}, {Krishnamurthy}  \& {Willett}}{{Borkowski}
  et~al.}{2013}]{Borkowski2013}
{Borkowski} K.~J.,  {Reynolds} S.~P.,  {Hwang} U.,  {Green} D.~A.,  {Petre} R.,
   {Krishnamurthy} K.,   {Willett} R.,  2013, \mn@doi [\apjl]
  {10.1088/2041-8205/771/1/L9}, \href
  {https://ui.adsabs.harvard.edu/abs/2013ApJ...771L...9B} {771, L9}

\bibitem[\protect\citeauthoryear{{Borkowski}, {Gwynne}, {Reynolds}, {Green},
  {Hwang}, {Petre}  \& {Willett}}{{Borkowski} et~al.}{2017}]{Borkowski2017}
{Borkowski} K.~J.,  {Gwynne} P.,  {Reynolds} S.~P.,  {Green} D.~A.,  {Hwang}
  U.,  {Petre} R.,   {Willett} R.,  2017, \mn@doi [\apjl]
  {10.3847/2041-8213/aa618c}, \href
  {https://ui.adsabs.harvard.edu/abs/2017ApJ...837L...7B} {837, L7}

\bibitem[\protect\citeauthoryear{{Brandner}, {Grebel}, {Chu}  \&
  {Weis}}{{Brandner} et~al.}{1997}]{Brandner1997}
{Brandner} W.,  {Grebel} E.~K.,  {Chu} Y.-H.,   {Weis} K.,  1997, \mn@doi
  [\apjl] {10.1086/310460}, \href
  {https://ui.adsabs.harvard.edu/abs/1997ApJ...475L..45B} {475, L45}

\bibitem[\protect\citeauthoryear{{Burkey}, {Reynolds}, {Borkowski}  \&
  {Blondin}}{{Burkey} et~al.}{2013}]{Burkey2013}
{Burkey} M.~T.,  {Reynolds} S.~P.,  {Borkowski} K.~J.,   {Blondin} J.~M.,
  2013, \mn@doi [\apj] {10.1088/0004-637X/764/1/63}, \href
  {https://ui.adsabs.harvard.edu/abs/2013ApJ...764...63B} {764, 63}

\bibitem[\protect\citeauthoryear{{Castelletti}, {Dubner}, {Golap}  \&
  {Goss}}{{Castelletti} et~al.}{2006}]{Castelletti2006}
{Castelletti} G.,  {Dubner} G.,  {Golap} K.,   {Goss} W.~M.,  2006, \mn@doi
  [\aap] {10.1051/0004-6361:20065061}, \href
  {https://ui.adsabs.harvard.edu/abs/2006A&A...459..535C} {459, 535}

\bibitem[\protect\citeauthoryear{{Chevalier}}{{Chevalier}}{1982a}]{Chevalier82}
{Chevalier} R.~A.,  1982a, \mn@doi [\apj] {10.1086/160126}, \href
  {https://ui.adsabs.harvard.edu/abs/1982ApJ...258..790C} {258, 790}

\bibitem[\protect\citeauthoryear{{Chevalier}}{{Chevalier}}{1982b}]{Chevalier1982b}
{Chevalier} R.~A.,  1982b, \mn@doi [\apj] {10.1086/160167}, \href
  {https://ui.adsabs.harvard.edu/abs/1982ApJ...259..302C} {259, 302}

\bibitem[\protect\citeauthoryear{{Chevalier}, {Blondin}  \&
  {Emmering}}{{Chevalier} et~al.}{1992}]{Chevalier1992}
{Chevalier} R.~A.,  {Blondin} J.~M.,   {Emmering} R.~T.,  1992, \mn@doi [\apj]
  {10.1086/171411}, \href
  {https://ui.adsabs.harvard.edu/abs/1992ApJ...392..118C} {392, 118}

\bibitem[\protect\citeauthoryear{{Chiotellis}, {Schure}  \&
  {Vink}}{{Chiotellis} et~al.}{2012}]{Chiotellis12}
{Chiotellis} A.,  {Schure} K.~M.,   {Vink} J.,  2012, \mn@doi [\aap]
  {10.1051/0004-6361/201014754}, \href
  {http://adsabs.harvard.edu/abs/2012A%26A...537A.139C} {537, A139}

\bibitem[\protect\citeauthoryear{{Chiotellis}, {Boumis}  \&
  {Spetsieri}}{{Chiotellis} et~al.}{2020}]{Chiotellis2020}
{Chiotellis} A.,  {Boumis} P.,   {Spetsieri} Z.~T.,  2020, \mn@doi [Galaxies]
  {10.3390/galaxies8020038}, \href
  {https://ui.adsabs.harvard.edu/abs/2020Galax...8...38C} {8, 38}

\bibitem[\protect\citeauthoryear{{Chugai}}{{Chugai}}{2001}]{Chugai2001}
{Chugai} N.~N.,  2001, \mn@doi [\mnras] {10.1111/j.1365-2966.2001.04717.x},
  \href {https://ui.adsabs.harvard.edu/abs/2001MNRAS.326.1448C} {326, 1448}

\bibitem[\protect\citeauthoryear{{Cikota}, {Patat}, {Cikota}, {Spyromilio}  \&
  {Rau}}{{Cikota} et~al.}{2017}]{Cikota2017}
{Cikota} A.,  {Patat} F.,  {Cikota} S.,  {Spyromilio} J.,   {Rau} G.,  2017,
  \mn@doi [\mnras] {10.1093/mnras/stx1734}, \href
  {https://ui.adsabs.harvard.edu/abs/2017MNRAS.471.2111C} {471, 2111}

\bibitem[\protect\citeauthoryear{{Clark}, {L{\'o}pez}, {Edwards}  \&
  {Winge}}{{Clark} et~al.}{2014}]{Clark2014}
{Clark} D.~M.,  {L{\'o}pez} J.~A.,  {Edwards} M.~L.,   {Winge} C.,  2014,
  \mn@doi [\aj] {10.1088/0004-6256/148/5/98}, \href
  {https://ui.adsabs.harvard.edu/abs/2014AJ....148...98C} {148, 98}

\bibitem[\protect\citeauthoryear{{Clyne}, {Akras}, {Steffen}, {Redman},
  {Gon{\c{c}}alves}  \& {Harvey}}{{Clyne} et~al.}{2015}]{Clyne2015}
{Clyne} N.,  {Akras} S.,  {Steffen} W.,  {Redman} M.~P.,  {Gon{\c{c}}alves}
  D.~R.,   {Harvey} E.,  2015, \mn@doi [\aap] {10.1051/0004-6361/201526585},
  \href {https://ui.adsabs.harvard.edu/abs/2015A&A...582A..60C} {582, A60}

\bibitem[\protect\citeauthoryear{{Crotts} \& {Heathcote}}{{Crotts} \&
  {Heathcote}}{1991}]{Crotts1991}
{Crotts} A.~P.,  {Heathcote} S.~R.,  1991, \mn@doi [\nat] {10.1038/350683a0},
  \href {https://ui.adsabs.harvard.edu/abs/1991Natur.350..683C} {350, 683}

\bibitem[\protect\citeauthoryear{{Decin} et~al.,}{{Decin}
  et~al.}{2020}]{Decin2020S}
{Decin} L.,  et~al., 2020, \mn@doi [Science] {10.1126/science.abb1229}, \href
  {https://ui.adsabs.harvard.edu/abs/2020Sci...369.1497D} {369, 1497}

\bibitem[\protect\citeauthoryear{{Drew} et~al.,}{{Drew}
  et~al.}{2005}]{Drew2005}
{Drew} J.~E.,  et~al., 2005, \mn@doi [\mnras]
  {10.1111/j.1365-2966.2005.09330.x}, \href
  {https://ui.adsabs.harvard.edu/abs/2005MNRAS.362..753D} {362, 753}

\bibitem[\protect\citeauthoryear{{Dwarkadas} \& {Chevalier}}{{Dwarkadas} \&
  {Chevalier}}{1998}]{Dwarkadas1998}
{Dwarkadas} V.~V.,  {Chevalier} R.~A.,  1998, \mn@doi [\apj] {10.1086/305478},
  \href {https://ui.adsabs.harvard.edu/abs/1998ApJ...497..807D} {497, 807}

\bibitem[\protect\citeauthoryear{{Ellison}, {Decourchelle}  \&
  {Ballet}}{{Ellison} et~al.}{2004}]{Ellison2004}
{Ellison} D.~C.,  {Decourchelle} A.,   {Ballet} J.,  2004, \mn@doi [\aap]
  {10.1051/0004-6361:20034073}, \href
  {https://ui.adsabs.harvard.edu/abs/2004A&A...413..189E} {413, 189}

\bibitem[\protect\citeauthoryear{{Fabrika}}{{Fabrika}}{2004}]{Fabrika2004}
{Fabrika} S.,  2004, Astrophys. Space Phys. Rev.,, \href
  {https://ui.adsabs.harvard.edu/abs/2004ASPRv..12....1F} {12, 1}

\bibitem[\protect\citeauthoryear{{Frew}, {Parker}  \&
  {Boji{\v{c}}i{\'c}}}{{Frew} et~al.}{2016}]{Frew2016}
{Frew} D.~J.,  {Parker} Q.~A.,   {Boji{\v{c}}i{\'c}} I.~S.,  2016, \mn@doi
  [\mnras] {10.1093/mnras/stv1516}, \href
  {https://ui.adsabs.harvard.edu/abs/2016MNRAS.455.1459F} {455, 1459}

\bibitem[\protect\citeauthoryear{{Gaensler}, {Green}  \&
  {Manchester}}{{Gaensler} et~al.}{1998}]{Gaensler1998}
{Gaensler} B.~M.,  {Green} A.~J.,   {Manchester} R.~N.,  1998, \mn@doi [\mnras]
  {10.1046/j.1365-8711.1998.01814.x}, \href
  {https://ui.adsabs.harvard.edu/abs/1998MNRAS.299..812G} {299, 812}

\bibitem[\protect\citeauthoryear{{Garc{\'\i}a-Segura}, {Langer},
  {R{\'o}{\.z}yczka}  \& {Franco}}{{Garc{\'\i}a-Segura}
  et~al.}{1999}]{GarciaSegura1999}
{Garc{\'\i}a-Segura} G.,  {Langer} N.,  {R{\'o}{\.z}yczka} M.,   {Franco} J.,
  1999, \mn@doi [\apj] {10.1086/307205}, \href
  {https://ui.adsabs.harvard.edu/abs/1999ApJ...517..767G} {517, 767}

\bibitem[\protect\citeauthoryear{{Grichener} \& {Soker}}{{Grichener} \&
  {Soker}}{2017}]{Grichener2017}
{Grichener} A.,  {Soker} N.,  2017, \mn@doi [\mnras] {10.1093/mnras/stx534},
  \href {https://ui.adsabs.harvard.edu/abs/2017MNRAS.468.1226G} {468, 1226}

\bibitem[\protect\citeauthoryear{{Hamuy} et~al.,}{{Hamuy}
  et~al.}{2003}]{Hamuy2003}
{Hamuy} M.,  et~al., 2003, \mn@doi [\nat] {10.1038/nature01854}, \href
  {https://ui.adsabs.harvard.edu/abs/2003Natur.424..651H} {424, 651}

\bibitem[\protect\citeauthoryear{{Heger}, {Langer}  \& {Woosley}}{{Heger}
  et~al.}{2000}]{Heger2000}
{Heger} A.,  {Langer} N.,   {Woosley} S.~E.,  2000, \mn@doi [\apj]
  {10.1086/308158}, \href
  {https://ui.adsabs.harvard.edu/abs/2000ApJ...528..368H} {528, 368}

\bibitem[\protect\citeauthoryear{{Hillebrandt}, {Hoeflich}, {Weiss}  \&
  {Truran}}{{Hillebrandt} et~al.}{1987}]{Hillebrandt1987}
{Hillebrandt} W.,  {Hoeflich} P.,  {Weiss} A.,   {Truran} J.~W.,  1987, \mn@doi
  [\nat] {10.1038/327597a0}, \href
  {https://ui.adsabs.harvard.edu/abs/1987Natur.327..597H} {327, 597}

\bibitem[\protect\citeauthoryear{{Huggins}, {Mauron}  \& {Wirth}}{{Huggins}
  et~al.}{2009}]{Huggins2009}
{Huggins} P.~J.,  {Mauron} N.,   {Wirth} E.~A.,  2009, \mn@doi [\mnras]
  {10.1111/j.1365-2966.2009.14874.x}, \href
  {https://ui.adsabs.harvard.edu/abs/2009MNRAS.396.1805H} {396, 1805}

\bibitem[\protect\citeauthoryear{{Igumenshchev}, {Tutukov}  \&
  {Shustov}}{{Igumenshchev} et~al.}{1992}]{Igumenshchev1992}
{Igumenshchev} I.~V.,  {Tutukov} A.~V.,   {Shustov} B.~M.,  1992, \sovast,
  \href {https://ui.adsabs.harvard.edu/abs/1992SvA....36..241I} {36, 241}

\bibitem[\protect\citeauthoryear{{Katsuda} et~al.,}{{Katsuda}
  et~al.}{2016}]{Katsuda2016}
{Katsuda} S.,  et~al., 2016, \mn@doi [\apj] {10.3847/0004-637X/832/2/194},
  \href {https://ui.adsabs.harvard.edu/abs/2016ApJ...832..194K} {832, 194}

\bibitem[\protect\citeauthoryear{{Keppens}, {Nool}, {T{\'o}th}  \&
  {Goedbloed}}{{Keppens} et~al.}{2003}]{Keppens03}
{Keppens} R.,  {Nool} M.,  {T{\'o}th} G.,   {Goedbloed} J.~P.,  2003, \mn@doi
  [Computer Physics Communications] {10.1016/S0010-4655(03)00139-5}, \href
  {http://adsabs.harvard.edu/abs/2003CoPhC.153..317K} {153, 317}

\bibitem[\protect\citeauthoryear{{Kramer}, {Lyne}, {Hobbs}, {L{\"o}hmer},
  {Carr}, {Jordan}  \& {Wolszczan}}{{Kramer} et~al.}{2003}]{Kramer2003}
{Kramer} M.,  {Lyne} A.~G.,  {Hobbs} G.,  {L{\"o}hmer} O.,  {Carr} P.,
  {Jordan} C.,   {Wolszczan} A.,  2003, \mn@doi [\apjl] {10.1086/378082}, \href
  {https://ui.adsabs.harvard.edu/abs/2003ApJ...593L..31K} {593, L31}

\bibitem[\protect\citeauthoryear{{Livio} \& {Mazzali}}{{Livio} \&
  {Mazzali}}{2018}]{Livio2018}
{Livio} M.,  {Mazzali} P.,  2018, \mn@doi [\physrep]
  {10.1016/j.physrep.2018.02.002}, \href
  {https://ui.adsabs.harvard.edu/abs/2018PhR...736....1L} {736, 1}

\bibitem[\protect\citeauthoryear{{Mastrodemos} \& {Morris}}{{Mastrodemos} \&
  {Morris}}{1999}]{Mastrodemos1999}
{Mastrodemos} N.,  {Morris} M.,  1999, \mn@doi [\apj] {10.1086/307717}, \href
  {https://ui.adsabs.harvard.edu/abs/1999ApJ...523..357M} {523, 357}

\bibitem[\protect\citeauthoryear{{Matt} \& {Balick}}{{Matt} \&
  {Balick}}{2004}]{Matt2004}
{Matt} S.,  {Balick} B.,  2004, \mn@doi [\apj] {10.1086/424727}, \href
  {https://ui.adsabs.harvard.edu/abs/2004ApJ...615..921M} {615, 921}

\bibitem[\protect\citeauthoryear{{Mauerhan} et~al.,}{{Mauerhan}
  et~al.}{2014}]{Mauerhan2014}
{Mauerhan} J.,  et~al., 2014, \mn@doi [\mnras] {10.1093/mnras/stu730}, \href
  {https://ui.adsabs.harvard.edu/abs/2014MNRAS.442.1166M} {442, 1166}

\bibitem[\protect\citeauthoryear{{Meaburn}}{{Meaburn}}{1987}]{Meaburn1987}
{Meaburn} J.,  1987, \mn@doi [\mnras] {10.1093/mnras/229.3.457}, \href
  {https://ui.adsabs.harvard.edu/abs/1987MNRAS.229..457M} {229, 457}

\bibitem[\protect\citeauthoryear{{Millas}, {Porth}  \& {Keppens}}{{Millas}
  et~al.}{2019}]{Millas2019}
{Millas} D.,  {Porth} O.,   {Keppens} R.,  2019, in {Sauty} C.,  ed.,  Vol. 55,
  JET Simulations, Experiments, and Theory: Ten Years After JETSET. What Is
  Next?. p.~71, \mn@doi{10.1007/978-3-030-14128-8_11}

\bibitem[\protect\citeauthoryear{{O'Connor}, {Redman}, {Holloway}, {Bryce},
  {L{\'o}pez}  \& {Meaburn}}{{O'Connor} et~al.}{2000}]{Conor2000}
{O'Connor} J.~A.,  {Redman} M.~P.,  {Holloway} A.~J.,  {Bryce} M.,  {L{\'o}pez}
  J.~A.,   {Meaburn} J.,  2000, \mn@doi [\apj] {10.1086/308452}, \href
  {https://ui.adsabs.harvard.edu/abs/2000ApJ...531..336O} {531, 336}

\bibitem[\protect\citeauthoryear{{Orlando} et~al.,}{{Orlando}
  et~al.}{2020}]{Orlando2020}
{Orlando} S.,  et~al., 2020, \mn@doi [\aap] {10.1051/0004-6361/201936718},
  \href {https://ui.adsabs.harvard.edu/abs/2020A&A...636A..22O} {636, A22}

\bibitem[\protect\citeauthoryear{{Plait}, {Lundqvist}, {Chevalier}  \&
  {Kirshner}}{{Plait} et~al.}{1995}]{Plait1995}
{Plait} P.~C.,  {Lundqvist} P.,  {Chevalier} R.~A.,   {Kirshner} R.~P.,  1995,
  \mn@doi [\apj] {10.1086/175213}, \href
  {https://ui.adsabs.harvard.edu/abs/1995ApJ...439..730P} {439, 730}

\bibitem[\protect\citeauthoryear{{Politano} \& {Taam}}{{Politano} \&
  {Taam}}{2011}]{Politano2011}
{Politano} M.,  {Taam} R.~E.,  2011, \mn@doi [\apj]
  {10.1088/0004-637X/741/1/5}, \href
  {https://ui.adsabs.harvard.edu/abs/2011ApJ...741....5P} {741, 5}

\bibitem[\protect\citeauthoryear{{Reilly}, {Maund}, {Baade}, {Wheeler},
  {H{\"o}flich}, {Spyromilio}, {Patat}  \& {Wang}}{{Reilly}
  et~al.}{2017}]{Reilly2017}
{Reilly} E.,  {Maund} J.~R.,  {Baade} D.,  {Wheeler} J.~C.,  {H{\"o}flich} P.,
  {Spyromilio} J.,  {Patat} F.,   {Wang} L.,  2017, \mn@doi [\mnras]
  {10.1093/mnras/stx1228}, \href
  {https://ui.adsabs.harvard.edu/abs/2017MNRAS.470.1491R} {470, 1491}

\bibitem[\protect\citeauthoryear{{Reynolds}, {Borkowski}, {Hwang}, {Hughes},
  {Badenes}, {Laming}  \& {Blondin}}{{Reynolds} et~al.}{2007}]{Reynolds2007}
{Reynolds} S.~P.,  {Borkowski} K.~J.,  {Hwang} U.,  {Hughes} J.~P.,  {Badenes}
  C.,  {Laming} J.~M.,   {Blondin} J.~M.,  2007, \mn@doi [\apjl]
  {10.1086/522830}, \href
  {https://ui.adsabs.harvard.edu/abs/2007ApJ...668L.135R} {668, L135}

\bibitem[\protect\citeauthoryear{{Santander-Garc{\'\i}a}, {Rodr{\'\i}guez-Gil},
  {Corradi}, {Jones}, {Miszalski}, {Boffin}, {Rubio-D{\'\i}ez}  \&
  {Kotze}}{{Santander-Garc{\'\i}a} et~al.}{2015}]{Santander2015}
{Santander-Garc{\'\i}a} M.,  {Rodr{\'\i}guez-Gil} P.,  {Corradi} R.~L.~M.,
  {Jones} D.,  {Miszalski} B.,  {Boffin} H.~M.~J.,  {Rubio-D{\'\i}ez} M.~M.,
  {Kotze} M.~M.,  2015, \mn@doi [\nat] {10.1038/nature14124}, \href
  {https://ui.adsabs.harvard.edu/abs/2015Natur.519...63S} {519, 63}

\bibitem[\protect\citeauthoryear{{Schure}, {Kosenko}, {Kaastra}, {Keppens}  \&
  {Vink}}{{Schure} et~al.}{2009}]{Schure09}
{Schure} K.~M.,  {Kosenko} D.,  {Kaastra} J.~S.,  {Keppens} R.,   {Vink} J.,
  2009, \mn@doi [\aap] {10.1051/0004-6361/200912495}, \href
  {http://adsabs.harvard.edu/abs/2009A%26A...508..751S} {508, 751}

\bibitem[\protect\citeauthoryear{{Smith}}{{Smith}}{2002}]{Smith2002}
{Smith} N.,  2002, \mn@doi [\mnras] {10.1046/j.1365-8711.2002.05966.x}, \href
  {https://ui.adsabs.harvard.edu/abs/2002MNRAS.337.1252S} {337, 1252}

\bibitem[\protect\citeauthoryear{{Smith}}{{Smith}}{2007}]{Smith2007b}
{Smith} N.,  2007, \mn@doi [\aj] {10.1086/510838}, \href
  {https://ui.adsabs.harvard.edu/abs/2007AJ....133.1034S} {133, 1034}

\bibitem[\protect\citeauthoryear{{Smith} \& {Arnett}}{{Smith} \&
  {Arnett}}{2014}]{Smith2014}
{Smith} N.,  {Arnett} W.~D.,  2014, \mn@doi [\apj]
  {10.1088/0004-637X/785/2/82}, \href
  {https://ui.adsabs.harvard.edu/abs/2014ApJ...785...82S} {785, 82}

\bibitem[\protect\citeauthoryear{{Smith}, {Bally}  \& {Walawender}}{{Smith}
  et~al.}{2007}]{Smith2007}
{Smith} N.,  {Bally} J.,   {Walawender} J.,  2007, \mn@doi [\aj]
  {10.1086/518563}, \href
  {https://ui.adsabs.harvard.edu/abs/2007AJ....134..846S} {134, 846}

\bibitem[\protect\citeauthoryear{{Soker}}{{Soker}}{2019}]{Soker2019}
{Soker} N.,  2019, \mn@doi [\nar] {10.1016/j.newar.2020.101535}, \href
  {https://ui.adsabs.harvard.edu/abs/2019NewAR..8701535S} {87, 101535}

\bibitem[\protect\citeauthoryear{{Tiffany}, {Humphreys}, {Jones}  \&
  {Davidson}}{{Tiffany} et~al.}{2010}]{Tiffany2010}
{Tiffany} C.,  {Humphreys} R.~M.,  {Jones} T.~J.,   {Davidson} K.,  2010,
  \mn@doi [\aj] {10.1088/0004-6256/140/2/339}, \href
  {https://ui.adsabs.harvard.edu/abs/2010AJ....140..339T} {140, 339}

\bibitem[\protect\citeauthoryear{{Townsend} \& {Owocki}}{{Townsend} \&
  {Owocki}}{2005}]{Townsend2005}
{Townsend} R.~H.~D.,  {Owocki} S.~P.,  2005, \mn@doi [\mnras]
  {10.1111/j.1365-2966.2005.08642.x}, \href
  {https://ui.adsabs.harvard.edu/abs/2005MNRAS.357..251T} {357, 251}

\bibitem[\protect\citeauthoryear{{Truelove} \& {McKee}}{{Truelove} \&
  {McKee}}{1999}]{Truelove99}
{Truelove} J.~K.,  {McKee} C.~F.,  1999, \mn@doi [\apjs] {10.1086/313176},
  \href {http://adsabs.harvard.edu/abs/1999ApJS..120..299T} {120, 299}

\bibitem[\protect\citeauthoryear{{Tsebrenko} \& {Soker}}{{Tsebrenko} \&
  {Soker}}{2013}]{Tsebrenko2013}
{Tsebrenko} D.,  {Soker} N.,  2013, \mn@doi [\mnras] {10.1093/mnras/stt1301},
  \href {https://ui.adsabs.harvard.edu/abs/2013MNRAS.435..320T} {435, 320}

\bibitem[\protect\citeauthoryear{{Tsebrenko} \& {Soker}}{{Tsebrenko} \&
  {Soker}}{2015a}]{Tsebrenko2015b}
{Tsebrenko} D.,  {Soker} N.,  2015a, \mn@doi [\mnras] {10.1093/mnras/stu2567},
  \href {https://ui.adsabs.harvard.edu/abs/2015MNRAS.447.2568T} {447, 2568}

\bibitem[\protect\citeauthoryear{{Tsebrenko} \& {Soker}}{{Tsebrenko} \&
  {Soker}}{2015b}]{Tsebrenko2015}
{Tsebrenko} D.,  {Soker} N.,  2015b, \mn@doi [\mnras] {10.1093/mnras/stv669},
  \href {https://ui.adsabs.harvard.edu/abs/2015MNRAS.450.1399T} {450, 1399}

\bibitem[\protect\citeauthoryear{{Tsebrenko} \& {Soker}}{{Tsebrenko} \&
  {Soker}}{2015c}]{Tsebrenko2015c}
{Tsebrenko} D.,  {Soker} N.,  2015c, \mn@doi [\mnras] {10.1093/mnras/stv1641},
  \href {https://ui.adsabs.harvard.edu/abs/2015MNRAS.453..166T} {453, 166}

\bibitem[\protect\citeauthoryear{{Vink}}{{Vink}}{2008}]{Vink2008}
{Vink} J.,  2008, \mn@doi [\apj] {10.1086/592375}, \href
  {https://ui.adsabs.harvard.edu/abs/2008ApJ...689..231V} {689, 231}

\bibitem[\protect\citeauthoryear{{Warren} \& {Blondin}}{{Warren} \&
  {Blondin}}{2013}]{Warren2013}
{Warren} D.~C.,  {Blondin} J.~M.,  2013, \mn@doi [\mnras]
  {10.1093/mnras/sts566}, \href
  {https://ui.adsabs.harvard.edu/abs/2013MNRAS.429.3099W} {429, 3099}

\bibitem[\protect\citeauthoryear{{Warren}, {Hughes}  \& {Slane}}{{Warren}
  et~al.}{2003}]{Warren2003}
{Warren} J.~S.,  {Hughes} J.~P.,   {Slane} P.~O.,  2003, \mn@doi [\apj]
  {10.1086/345078}, \href
  {https://ui.adsabs.harvard.edu/abs/2003ApJ...583..260W} {583, 260}

\bibitem[\protect\citeauthoryear{{Weaver}, {McCray}, {Castor}, {Shapiro}  \&
  {Moore}}{{Weaver} et~al.}{1977}]{Weaver1977}
{Weaver} R.,  {McCray} R.,  {Castor} J.,  {Shapiro} P.,   {Moore} R.,  1977,
  \mn@doi [\apj] {10.1086/155692}, \href
  {https://ui.adsabs.harvard.edu/abs/1977ApJ...218..377W} {218, 377}

\bibitem[\protect\citeauthoryear{{West}, {Lauberts}, {Jorgensen}  \&
  {Schuster}}{{West} et~al.}{1987}]{West1987}
{West} R.~M.,  {Lauberts} A.,  {Jorgensen} H.~E.,   {Schuster} H.~E.,  1987,
  \aap, \href {https://ui.adsabs.harvard.edu/abs/1987A&A...177L...1W} {177, L1}

\bibitem[\protect\citeauthoryear{{Williams}, {Borkowski}, {Reynolds},
  {Ghavamian}, {Blair}, {Long}  \& {Sankrit}}{{Williams}
  et~al.}{2012}]{Williams2012}
{Williams} B.~J.,  {Borkowski} K.~J.,  {Reynolds} S.~P.,  {Ghavamian} P.,
  {Blair} W.~P.,  {Long} K.~S.,   {Sankrit} R.,  2012, \mn@doi [\apj]
  {10.1088/0004-637X/755/1/3}, \href
  {https://ui.adsabs.harvard.edu/abs/2012ApJ...755....3W} {755, 3}

\bibitem[\protect\citeauthoryear{{Yu} \& {Fang}}{{Yu} \& {Fang}}{2018}]{Yu2018}
{Yu} H.,  {Fang} J.,  2018, \mn@doi [Research in Astronomy and Astrophysics]
  {10.1088/1674-4527/18/9/117}, \href
  {https://ui.adsabs.harvard.edu/abs/2018RAA....18..117Y} {18, 117}

\makeatother
\end{thebibliography}

\end{document}